\documentclass{article}

\usepackage{PRIMEarxiv}

\usepackage[utf8]{inputenc} % allow utf-8 input
\usepackage[T1]{fontenc}    % use 8-bit T1 fonts
\usepackage{hyperref}       % hyperlinks
\usepackage{url}            % simple URL typesetting
\usepackage{booktabs}       % professional-quality tables
\usepackage{amsfonts}       % blackboard math symbols
\usepackage{nicefrac}       % compact symbols for 1/2, etc.
\usepackage{microtype}      % microtypography
\usepackage{lipsum}
\usepackage{fancyhdr}       % header
\usepackage{graphicx}       % graphics
\graphicspath{{media/}}     % organize your images and other figures under media/ folder

\usepackage{color}
\usepackage{amsmath, amssymb, amsthm, bm}
\usepackage{subfig}
\usepackage{comment}
\usepackage{multirow}

\newcommand\norm[1]{\left\lVert #1 \right\rVert}

\newtheorem{definition}{Definition}

\newtheorem{assumption}{Assumption}

\title{Driving Towards Stability and Efficiency: A Variable Time Gap Strategy for Adaptive Cruise Control
}

\author{
  Shaimaa K.~El-Baklish \\
  Institute for Transport Planning and Systems \\
  Department of Civil, Environmental and Geomatic Engineering \\ 
  ETH Zurich \\
  Zurich, Switzerland\\
  \texttt{selbaklish@ethz.ch} \\
  \AND
  Anastasios Kouvelas \\
  Institute for Transport Planning and Systems \\
  Department of Civil, Environmental and Geomatic Engineering \\ 
  ETH Zurich \\
  Zurich, Switzerland\\
  \texttt{kouvelas@ethz.ch} \\
  \AND
  Michail A.~Makridis \\
  Institute for Transport Planning and Systems \\
  Department of Civil, Environmental and Geomatic Engineering \\ 
  ETH Zurich \\
  Zurich, Switzerland\\
  \texttt{mmakridis@ethz.ch} \\
}

\begin{document}
\maketitle

\begin{abstract}

Automated vehicle technologies offer a promising avenue for enhancing traffic efficiency, safety, and energy consumption. Among these, Adaptive Cruise Control (ACC) systems stand out as a prevalent form of automation on today's roads, with their time gap settings holding paramount importance. While decreasing the average time headway tends to enhance traffic capacity, it simultaneously raises concerns regarding safety and string stability. This study introduces a novel variable time gap feedback control policy aimed at striking a balance between maintaining a minimum time gap setting under equilibrium car-following conditions, thereby improving traffic capacity, while ensuring string stability to mitigate disturbances away from the equilibrium flow. Leveraging nonlinear $\mathcal{H}_\infty$ control technique, the strategy employs a variable time gap component as the manipulated control signal, complemented by a constant time gap component that predominates during car-following equilibrium. The effectiveness of the proposed scheme is evaluated against its constant time-gap counterpart calibrated using field platoon data from the OpenACC dataset. Through numerical and traffic simulations, our findings illustrate that the proposed algorithm effectively dampens perturbations within vehicle platoons, leading to a more efficient and safer mixed traffic flow.
% 184 words

%In vehicle platooning, time gap settings of Adaptive Cruise Control (ACC) systems have a significant impact on car-following dynamics, traffic capacity and road safety. Traffic capacity increases with the reduction of the average time headway; however, this raises concerns of safety and string stability. This work presents a variable time gap feedback control strategy to balance following a minimum time gap setting under equilibrium car-following conditions for increased traffic capacity; and guaranteeing string stability to attenuate disturbances away from the equilibrium flow. This is achieved using nonlinear $\mathcal{H}_\infty$ control; where a variable time gap component is set as the manipulated control signal. Also, a constant time gap component is present which dominates during car-following equilibrium and is prescribed to the minimum value. Numerical simulations demonstrate that the proposed scheme yields less perturbations in space headway compared to commercially available ACC systems from the OpenACC dataset; showcasing the potential benefits of better road utilization and increased capacity from a traffic perspective.

% 160 words
% In addition, the string stability of commercially available ACC-equipped vehicles is analyzed in a direct data-driven fashion; sidestepping the need for model calibration.
\end{abstract}

% keywords can be removed

\section{Introduction}
\label{SEC:Intro}
Adaptive Cruise Control (ACC) is a part of the advanced driver assistance systems (ADAS) and is designed specifically for vehicle longitudinal control. In fact, ACC systems are the most widespread Society of Engineers (SAE) level 1 automated vehcile (AV) technology in the current markets~\cite{sae_standard}. For instance, 16 out of 20 best-selling cars in the US market are equipped with an ACC system~\cite{Shang2021}. More importantly, vehicle automation gives rise to change traffic flow dynamics and to alleviate instabilities and congestion.

At its center, ACC employs a spacing policy which in turn determines the car-following behavior, platoon stability, and traffic efficiency~\cite{Wu2020}. A small inter-vehicular spacing results in a higher traffic capacity but undermines safety and stability. The notion of stability referred to in this work is string stability. Intuitively, vehicles in a platoon are string stable if small perturbations from an equilibrium flow are dampened as they propagate in the upstream direction. Spacing policies in the literature can be categorized into 1) constant spacing (CS), 2) constant time gap policies (CTG), and 3) variable time gap (VTG) policies~\cite{Ntousakis2015}. String stability was proved unattainable using CS policy without inter-vehicular communication to provide further information such as leader's acceleration~\cite{Ntousakis2015, Bian2018}. On the other hand, CTG and VTG policies can be appropriately designed to be string stable; nevertheless, there exists the conflicting objective of traffic efficiency. Using the OpenACC database for car-following experiments, Makridis et al.~confirm that a string-stable ACC with large time headway leads to poor road utilization decreasing capacity~\cite{Makridis2021}. In addition, a large headway threatens drivers' subjective acceptability due to possibly increased lane changes from adjacent lanes~\cite{Wu2020}.

%%%%%%%%%%%%%%%%%%%%%%%%%%%%%%%%%%%%%%%%%%%%%%%%%%%%%%%%%%%%%%%%%%%%%%%%%%%%%%%%%%%%%%%%%
%%%%%%%%%%%%%%%%%%%%%%%%%%%%%%%%%%%%%%%%%%%%%%%%%%%%%%%%%%%%%%%%%%%%%%%%%%%%%%%%%%%%%%%%%

\subsection{Related Work} \label{SEC:LitReview}
The most widely used ACC model is developed by Milanés and Shladover~\cite{Milanes2014} as a linear feedback control algorithm; where the acceleration is computed based on deviations away from a target speed (leading vehicle speed in car-following mode) and away from a target spacing determined by a constant time gap (CTG) policy. This spacing policy assumes that the space gap is proportional to the vehicle speed; thus, more appealing to drivers. Furthermore, Gunter et al.~have shown that commercially available ACC systems utilizing CTG policy are string unstable~\cite{Gunter2021}. Interestingly, Shang and Stern presented simulations of string-unstable ACC vehicles platoon with a minimum time gap which enhanced downstream capacity relative to their string-stable counterpart with a maximum time gap~\cite{Shang2021}. Lin et al.~demonstrated through experiments the need for different time gap settings in different driving scenarios to balance safety and driver's subjective acceptance~\cite{Lin2009}.

Other variable spacing policies assume a nonlinear relationship between the space gap and the vehicle speed; thus, improving stability properties and traffic capacity. Wang and Rajamani~proposed a nonlinear spacing policy as a function of speed and approaching rate; inspired by Greenshield's fundamental diagram (FD) and equations~\cite{Wang2002}. The approaching rate is the relative speed of the ego vehicle to the leading vehicle. To enforce such a spacing policy, a sliding mode controller is implemented where the spacing is considered as the sliding variable. They have reported a 44\% decrease in travel time compared to a CTG policy in simulations of a pipeline plus on-ramp network. In~\cite{Zhou2004}, the authors have developed a quadratic spacing policy as a function of vehicle speed based on their investigation of human drivers' spacing policy. They have designed the policy parameters to maximize the traffic capacity while imposing linear stability constraints. Again, a sliding mode controller is developed on top to implement the proposed spacing policy, and the control scheme was tested in simulations of a merging scenario. Compared to a modified Greenshields-inspired spacing policy suggested by Swaroop and Rajagopal~\cite{Swaroop1999}, the proposed scheme yielded a higher critical density and attenuated upstream perturbations in vehicles' accelerations. In addition, a data-driven linear $\mathcal{H}_\infty$ ACC controller is designed in~\cite{Zhao2024} using a CTG policy. The authors design an adaptive estimator for the unknown parameters such as the time lag of low-level vehicle dynamics. Simulations and analysis show the string stability of the designed data-driven scheme. All string stability analyses performed therein were in the strict sense and based on linear frequency-domain input-output $\mathcal{H}_\infty$ gains. For more details on string stability analysis methods, the reader is referred to~\cite{Feng2019} and~\cite{Montanino2021}.

Variable time gap (VTG) policies can also be designed to achieve variable spacing. In~\cite{Kesting2008}, an intelligent driving strategy layer was suggested to regulate the ACC parameter settings according to the perceived traffic state. The authors aggregated traffic ``situations'' into five traffic states --- free traffic, upstream jam front, congested traffic, downstream jam front, and bottleneck sections --- according to smoothed measurements of the space-mean speed and location information of active bottlenecks from the infrastructure. Then, a driving strategy matrix was developed to vary the ACC model parameters according to the perceived state. An Intelligent Driver Model (IDM) was used as a controller. For instance, the time gap was linearly decreased for the "downstream jam front" state. Simulations demonstrate a reduction in travel times for small penetration rates of 5\%. Another VTG policy was proposed by Wang et al.~\cite{Wang2014}; where the time gap was varied linearly with the ratio between the measured space gap and the threshold space gap. The threshold space gap discriminates between cruising and car-following modes. The proposed VTG policy incentivizes shorter time gaps at larger space gaps that indicate higher densities. A two-regime model predictive control (MPC) scheme was implemented on top; where the costs represented safety and cruising and car-following modes efficiency. Spiliopoulou et al.~proposed adapting the time gap in real-time to the traffic flow state at a motorway section~\cite{Spiliopoulou2018}. The applied time gap was the minimum of the desired setting of the driver and the suggested one that attempts to avoid congestion by increasing the static capacity of the section and, also, the discharge flow at active bottleneck locations. Microscopic simulation for a motorway stretch with an on-ramp bottleneck shows a decrease in average vehicle delay and fuel consumption. These works showed improved traffic flow, but, without guarantees on string stability.

The work by~\cite{Chen2019} adapted the nonlinear spacing policies by~\cite{Wang2002, Wang2004} based on macroscopic flow and~\cite{Yanakiev1998} based on approaching rate. Hence, the developed ACC time gap strategy combined the vehicle speed, approaching rate, preceding vehicle acceleration, free-flow speed, and jam density. Simulations using PTV-VISSIM of a single-lane highway with an on-ramp demonstrated diminished travel time and traffic delay compared to CTG policy and the VTG policy developed by~\cite{Yanakiev1998}. A comparison between CTG and VTG policies for ACC was conducted in~\cite{Dong2021}. The VTG policy employed varied with the approaching rate within the time gap bounds. The authors further investigated the linear stability and throughput of the mixed traffic flow. Both investigations and simulations have shown that a VTG policy is more likely to stabilize traffic flow and that an increased maximum time gap yields a more stable mixed traffic flow, but with a reduced capacity. Khound et al.~developed an adaptive time gap strategy to provide safer and more comfortable performance~\cite{Khound2023}. The authors have shown through stability analysis the superiority of the adaptive time gap strategy compared to the CTG counterpart in the presence of actuator lag and intrinsic delays.
A macroscopic approach towards designing the VTG policy was introduced in~\cite{Bekiaris-Liberis2021}. The authors consider the time gap of ACC vehicles as a control input to be manipulated to stabilize traffic flow expressed by an Aw-Rascle-Zhang (ARZ) mixed traffic model. The feedback VTG control policy was developed for the linearized system around a uniform, congested equilibrium profile. Numerical simulations reported improved performance in terms of fuel consumption, travel time, and comfort compared to the CTG policy; which the authors proved to yield an unstable traffic flow. These works highlight the importance of the operating time gap strategy and its impact on traffic flow efficiency. 

%%%%%%%%%%%%%%%%%%%%%%%%%%%%%%%%%%%%%%%%%%%%%%%%%%%%%%%%%%%%%%%%%%%%%%%%%%%%%%%%%%%%%%%%%
%%%%%%%%%%%%%%%%%%%%%%%%%%%%%%%%%%%%%%%%%%%%%%%%%%%%%%%%%%%%%%%%%%%%%%%%%%%%%%%%%%%%%%%%%
\subsection{Objectives and Contributions}

\begin{figure}[!h]
    \centering
    \includegraphics[width=0.75\textwidth]{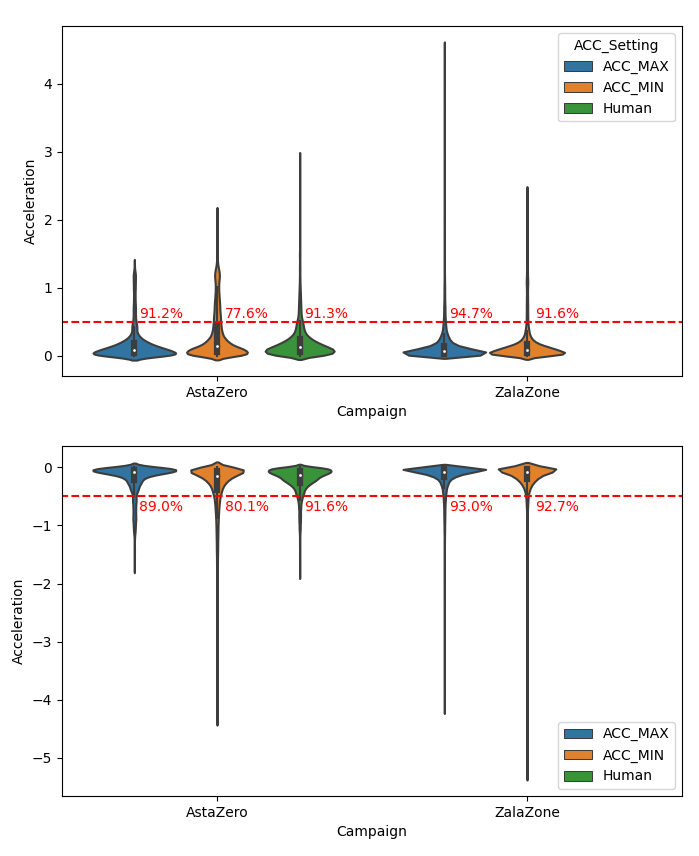}
    \caption{Distributions of instantaneous acceleration and deceleration of vehicles per campaign and driving mode. The red labels denote the cumulative proportion of acceleration/deceleration that is below/above $\pm 0.5 \text{ m/s}^2$.}
    \label{fig:Dist_Accel_Decel}
\end{figure}

This work aims to develop a VTG policy for an ACC controller that could achieve a balance between string stability and traffic efficiency. Under equilibrium flow, a CTG policy with a minimum time gap is adopted to promote efficient road utilization by ACC vehicles. This relies on the fact that stable driving conditions are most prevalent. This can also be supported by observed trajectory data from the OpenACC database. Figure~\ref{fig:Dist_Accel_Decel} shows the distributions of instantaneous acceleration and deceleration which are mostly under $\pm 0.5 \text{ m/s}^2$ for all platoons in the AstaZero and ZalaZone campaigns. However, perturbations/disturbances give rise to sharp acceleration/deceleration; raising safety, stability, and efficiency concerns. Further investigations by Ciuffo et.al.\ compared short, medium and large time gap settings for field platoons in the ZalaZone experimental campaign in terms of traffic capacity~\cite{Ciuffo2021}. They showed that short time gap platoons are able to attain the highest flow during stable conditions; while highly deteriorating during perturbations to relatively match medium time-gap platoons. Hence, string instability is detrimental to the potential benefits of ACC systems to traffic flow. Therefore, the CTG policy must be relaxed to allow for the needed larger gaps to attenuate such disturbances. 

The main contributions of this paper consist of the following aspects. First, a feedback $\mathcal{H}_\infty$-based VTG policy is adopted for guaranteed string stability; which consists of a constant and a manipulated component. The constant time gap component may be set to a desired setting by the driver or a minimum time gap for an efficient traffic flow. As for the manipulated time gap component, it is to be designed as a feedback control policy using nonlinear $\mathcal{H}_\infty$ synthesis to achieve $\mathcal{L}_2$ string stability in the strict sense. A nonlinear analysis is necessary here due to the VTG strategy adopted. Also, formulation through strict string stability facilitates extending the platoon to any number of vehicles; since a platoon can be split into multiple leader-follower subsystems. 

Our proposed methodology to design a reactive VTG strategy as a feedback control policy is similar to~\cite{Bekiaris-Liberis2021}. However, the proposed VTG strategy utilizes the space headway and velocity of the ego vehicle for state feedback; in contrast to traffic variables used by Bekiaris-Liberis and Delis~\cite{Bekiaris-Liberis2021}. Indeed, the microscopic states of the ego vehicle could be valid indicators for road utilization and, hence, traffic conditions. Therefore, the proposed VTG ACC controller is simpler to implement and may be able to balance traffic efficiency against string stability.

Furthermore, we analyze the robustness of the developed VTG ACC scheme against inherent uncertainties; specifically, the low-level vehicle lag dynamics and input time delay. Another important contribution presented here is a data-driven string stability analysis of vehicle platoons using empirical data collected from commercially available ACC-operated vehicles and manually driven vehicles. This approach is completely model-free and, hence, bypasses the need for calibration of car-following models to the experimental trajectory data to overcome any resulting biases. In addition, this allows for the performance assessment of platoon trajectories in terms of string stability without knowledge of the underlying ACC system operation.

The remainder of this paper is organized as follows. Platoon modeling and string stability are defined in Section~{\ref{SEC:Preliminaries}}. Then, we propose a data-driven string stability analysis approach for field platoons in Section~{\ref{SEC:L2gainEst}}. In Section~{\ref{SEC:ControlDesign}}, we design the $\mathcal{H}_\infty$-based VTG policy for adaptive cruise control. Finally, numerical simulations using field platoon data and traffic simulation using the microscopic simulator SUMO are discussed in Sections~{\ref{SEC:NumSim}} and~{\ref{SEC:SUMOSim}}, respectively, and we conclude thereafter in Section~{\ref{SEC:Conc}}.

\section{Preliminaries}
\label{SEC:Preliminaries}
Consider a platoon of $N+1$ vehicles. The leader is denoted by the index $0$. Each follower vehicle $i, \forall i = 1, \dots, N,$ is described by the following state-space model.
\begin{subequations}
\begin{align}
    & \dot{s}_i(t) = v_{i-1}(t) - v_i(t) \\
    & \dot{v}_i(t) = a_i(t) %f_{a,i}(s_i(t), v_i(t), v_{i-1}(t))
\end{align}
\label{EQ:general_CF_model}
\end{subequations}
where $s_i$ and $v_i$ denote the space headway and velocity of vehicle $i$ respectively and $v_{i-1}$ denotes the velocity of the preceding vehicle. An important distinction should be made between a model and a controller. To design an ACC controller, the acceleration command $a_i(t) \in \mathbb{R}$ is the control signal to be designed and is then executed through the low-level vehicle dynamics and actuators~\cite{Zhou2022_CommACCLowLevel}. On the other hand, car-following models are utilized for microscopic traffic simulations where $a_i(t) = f_{a,i}(s_i(t), v_i(t), v_{i-1}(t))$ in order to describe observed behaviors. The function $f_{a,i}(\cdot)$ represents the longitudinal dynamics defined by car-following models such as the IDM~\cite{Treiber2013}, which may be homogeneous or heterogeneous.

A signal $u(t)\in \mathcal{L}_2$ has, intuitively, bounded energy, i.e.\ $\int_{t_0}^{\infty} u^T(t) u(t) dt < \infty$, if it belongs to the Lebesgue space $\mathcal{L}_2$ of square-integrable functions.

\begin{definition}
\label{DEF:general_string_stability}
    Let $s_{\text{eq}, i}$ and $v_\text{eq}$ be the equilibrium profile of~\eqref{EQ:general_CF_model}, $\bm{x}_i(t) = \begin{bmatrix} s_i(t)-s_{\text{eq}, i} & v_i(t)-v_{\text{eq}, i} \end{bmatrix}^T$ be the deviation states away from the equilibrium profile and $\delta v_{i-1}(t) = v_{i-1} - v_\text{eq}$. For any platoon to be strictly input-to-state string-stable, follower vehicle $i$ should have a local $\mathcal{L}_2$-gain equal to or less than $\gamma, \gamma \leq 1$ for any initial state $\bm{x}_{i}(0) \in \mathcal{N}$ if the response $\bm{x}_i(t)$ to any disturbance from the preceding vehicle $\delta v_{i-1}(t) \in \mathcal{L}_2[0, \infty)$ satisfies
    \begin{equation}
        \norm{\bm{x}_i(t)}_{\mathcal{L}_2}^2 \leq \gamma^2 \Big( \kappa(\bm{x}_i(0)) + \norm{\delta v_{i-1}(t)}_{\mathcal{L}_2}^2  \Big), \forall t \geq 0, i= 1, \dots, N
    \end{equation}
    for some bounded function $\kappa$, $\kappa(0) = 0$ and $\mathcal{N} \subseteq \mathcal{X} \subseteq \mathbb{R}^2$ where $\mathcal{X}$ is the set of admissible states.
\end{definition}
\begin{definition}
\label{DEF:linear_string_stability}
    Let $\bm{X}_i(s)$ and $\Delta V_{i-1}(s)$ be the Laplace transform of $\bm{x}_i(t)$ and $\delta v_{i-1}(t)$. For a linear platoon to be strictly input-to-output string-stable, follower vehicle $i$ should have an $\mathcal{H}_\infty$-norm or induced $\mathcal{L}_2$-gain equal to or less than $\gamma, \gamma \leq 1$ if the response $\bm{x}_i(t)$ to any disturbance from the preceding vehicle $\delta v_{i-1}(t) \in \mathcal{L}_2[0, \infty)$ satisfies
    \begin{equation}
        \norm{\bm{\Gamma}_i(s)}_{\mathcal{H}_\infty} = \sup_{\bm{\delta v_{i-1}}\in\mathcal{L}_2(0, \infty)\neq 0} \frac{\norm{\bm{x}_i(t)}^2_{\mathcal{L}_2}}{\norm{\delta v_{i-1}(t)}^2_{\mathcal{L}_2}} \leq \gamma, \forall i= 1, \dots, N
    \end{equation}
    where $\bm{\Gamma_i(s)} = \frac{\bm{X}_i(s)}{\Delta V_{i-1}(s)}$ is the transfer function between follower vehicle $i$ and its predecessor $i-1$.
\end{definition}
The $\mathcal{L}_2$-gain, or $\mathcal{H}_\infty$-norm for linear systems, represents the worst-case system gain over all excitations or input disturbances. Thus, if a platoon has an $\mathcal{L}_2$-gain of $\gamma \leq 1$, it means that perturbations will dissipate as they propagate upstream and the platoon will be string-stable.

For a platoon of ACC vehicles~\cite{Milanes2014}, the dynamic model can be written as follows.
\begin{subequations}
\begin{align}
    & \dot{s}_i(t) = v_{i-1}(t) - v_i(t) \\
    & \dot{v}_i(t) = k_{1,i} (s_i(t) - s_0 - L_{i-1} - \tau_i v_i(t)) + k_{2,i} (v_{i-1}(t) - v_i(t))
\end{align}
\label{EQ:ACC_model_dyn}
\end{subequations}
where $s_0$ is the standstill distance, $L_{i-1}$ is the length of vehicle $i-1$ and $k_{1,i}, k_{2,i}$ and $\tau_i$ are the spacing control gain, velocity control gain and the constant time gap for ACC vehicle $i$. The CTG spacing policy is defined as $s_{\text{des}, i} = s_0 + L_{i-1} + \tau_i v_i(t)$.

To define the error dynamics of a leader-follower subsystem, the equilibrium profile is defined by the constant speed $v_\text{eq}$ at which all vehicles are traveling at. The equilibrium space headway is given as $s_{\text{eq}, i} = s_0 + L_{i-1} + \tau_i v_\text{eq}$. Then, the deviation states are expressed as $\tilde{s} _i= s_i - s_{\text{eq}, i}$ and $\tilde{v} = v_i - v_\text{eq}$ and the deviation dynamics are obtained as follows. 
\begin{subequations}
\begin{align}
    & \dot{\tilde{s}}_i = -\tilde{v}_i + \delta v_{i-1}
    \\
    & \dot{\tilde{v}}_i = k_{1,i} \tilde{s}_i - k_{1,i} \tau_i \tilde{v}_i - k_{2,i} \tilde{v}_i + k_{2,i} \delta v_{i-1}
\end{align}
\label{EQ:deviation_dyn}
\end{subequations}
where $\delta v_{i-1}$ denotes the deviation of the leader's velocity away from the equilibrium velocity and represents a disturbance to the considered leader-follower subsystem. The derivation can be found in \nameref{SEC:Appendix}.

%An important distinction should be made between a controller and a model. A controller

\section{Data-Driven String Stability Analysis of Adaptive Cruise Control}
\label{SEC:L2gainEst}
Due to propriety issues, the true model of ACC controllers is not known. Therefore, researchers proposed many models in the literature, most notably the one by~\cite{Milanes2014}, to analyze ACC controllers and their performance. However, retrieving a meaningful model that calibrates well to trajectory data from commercially available ACC systems is challenging requiring time and expert knowledge. Also, conclusions drawn from calibrated models may be subject to biases acquired from the optimization and calibration procedure. In this section, a direct data-driven approach for strict string stability analysis, with the assumption of a linear platoon model, is performed using only trajectory data and bypassing the need for model calibration~\cite{Kosut1995, VanHeusden2007}.

\subsection{Data-Driven $\mathcal{L}_2$-gain Estimation}
Consider a single-input single-output (SISO) linear time-invariant (LTI) system $G(s)$. In our case, the system input is the disturbing leader velocity $\delta v_{i-1}(t)$. The system output may be either the deviation of follower velocity $\tilde{v}_i(t)$ or the deviation of space headway $\tilde{s}_i(t)$. Trajectory data of length $N_D$ is assumed to be available. A matrix $T_m(\delta v_{i-1}) \in \mathbb{R}^{(N_D+m-1) \times m}$ is formed of $m$ columns of the downshifted input signal where $m < N_D$.
\begin{equation}
    T_m(\delta v_{i-1}) = \begin{bmatrix}
        \delta v_{i-1}(1) & 0 & \dots \\
        \delta v_{i-1}(2) & \delta v_{i-1}(1) & \dots \\
        \vdots & \ddots & \ddots \\
        \delta v_{i-1}(N_D) & \vdots & \ddots  \\
        0 & \delta v_{i-1}(N_D) & \ddots  \\
        \vdots & 0 & \ddots  \\
        \vdots & \vdots  & \ddots  \\
        
    \end{bmatrix}
\end{equation}
Then, the estimated auto-correlation matrix is defined as $\hat{R}_m(\delta v_{i-1}) = \frac{1}{N_D} T_m^T(\delta v_{i-1}) T_m(\delta v_{i-1}) \in \mathbb{R}^{m \times m}$. Each entry in this matrix represents estimates of the auto-correlation function $\hat{R}_{uu}(T) = \frac{1}{N_D}\sum_{t=1}^{N_D-T} u(t) u(t+T)$.

\begin{assumption}
    A linear time-invariant (LTI) system is persistently excited if the auto-correlation matrix of the input $u$ data trajectory of length $N_D$ is full rank, i.e.\ $\text{rank}(\hat{R}_m(u)) = \frac{1}{N_D} T_m^T(u) T_m(u) = m$.
    \label{ASSUM:persistent_excitation}
\end{assumption}

Similar matrices exist for the outputs as well. Provided that Assumption~\ref{ASSUM:persistent_excitation} is satisfied, we can define the following semi-definite program to estimate the $\mathcal{H}_\infty$-norm or $\mathcal{L}_2$-gain of the underlying system.
\begin{subequations}
\begin{align}
    \hat{\gamma} &= \min_{\gamma} \gamma \\
    & \text{s.t. } \frac{1}{N_D} T_m^T(y) T_m(y) - \gamma^2 \frac{1}{N_D} T_m^T(\delta v_{i-1}) T_m(\delta v_{i-1}) \leq 0 \text{ , } \gamma \geq 0
\end{align}
\label{EQ:data_driven_L2gain_estmation}
\end{subequations}
where $\hat{\gamma}$ is the estimated $\mathcal{L}_2$-gain and $y$ denotes the selected system output which can be $\tilde{v}_i(t)$ or $\tilde{s}_i(t)$. The estimated $\mathcal{L}_2$-gain converges to the true value, i.e.\ $\hat{\gamma}^2 \rightarrow \norm{G(s)}_{\mathcal{H}_\infty}^2$, as trajectory data length becomes infinite $N_D \rightarrow \infty$ $m \rightarrow \infty$ and $\frac{m}{N_D} \rightarrow 0$~\cite{Kosut1994}. This quantifies the system norm needed to make conclusions about strict string stability for linear platoons, as per Definition~\ref{DEF:linear_string_stability}.

\subsection{Field Data Description}
The empirical data were taken from two experimental campaigns in AstaZero and ZalaZone proving grounds in Sweden and Hungary respectively. The AstaZero campaign consisted of a fleet of 5 commercially available ACC-equipped vehicles on the 5.7-km-long Rural Road; while a fleet of 10 vehicles was deployed on the circular Dynamic Platform (300-m diameter) in the ZalaZone campaign. Both campaigns were carried out in 2019.

In both campaigns, data acquisition was performed using high-accuracy on-board equipment (e.g.\ U-blox global navigation satellite system (GNSS) receivers) with a sampling frequency of 10 Hz. Maximum and minimum time gap settings were used for the ACC-operated tests in both campaigns. Also, some tests were carried out without ACC operation (i.e.\ manual driving) in the AstaZero campaign. These data can be freely accessed through the OpenACC database~\cite{Makridis2021}.

\subsection{String Stability of Field Platoons}
Each platoon was divided into leader-follower subsystems; where the convex program~\eqref{EQ:data_driven_L2gain_estmation} was solved to estimate the $\mathcal{L}_2$-gain of the corresponding subsystem and, subsequently, provide numerical thresholds for strict string stability.

The speed-to-speed relation was examined; i.e.\ the output was taken to be the deviation in the follower velocity $y(t) = \tilde{v}_i (t)$. The equilibrium velocity $v_\text{eq}$ was taken to be the median velocity of the leading vehicle for 1-minute intervals in the duration of the car-following maneuver. 
%The resulting estimates were tested for and filtered from outliers using Grubbs test after testing the distribution of the estimated $\mathcal{L}_2$-gains to be normal using the Shapiro-Wilk test with a statistical significance level of $\alpha = 0.05$.

\begin{figure}[!ht]
    \centering
    \includegraphics[width=0.9\textwidth]{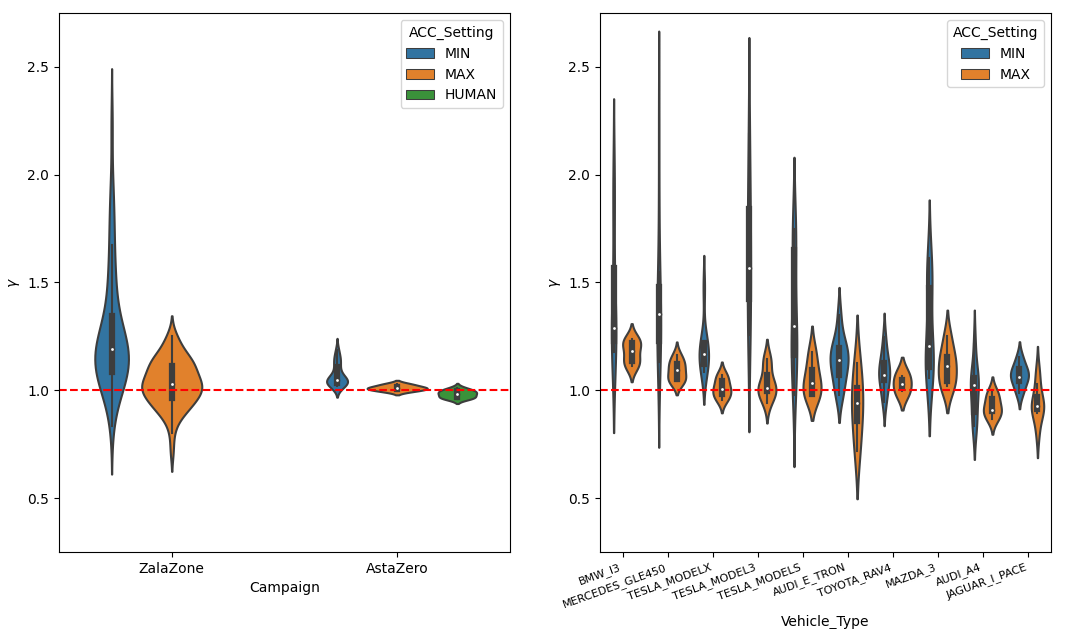}
    \caption{Distribution of estimated $\mathcal{L}_2$-gains for all leader-follower subsystems per campaign (left) and per vehicle make and model (right). The red dashed line represents the threshold of string stability at $\gamma = 1$.}
    \label{fig:L2gain_Summary_Campaign_VehicleType}
\end{figure}

Figure~\ref{fig:L2gain_Summary_Campaign_VehicleType} illustrates the distribution of the estimated $\mathcal{L}_2$-gain values of all leader-follower subsystems according to the ACC settings (``MIN''/``MAX'' for minimum/maximum time gap ACC setting and ``HUMAN'' for manual driving). A maximum time gap ACC setting has consistently lower $\mathcal{L}_2$-gains; even for the same make and model of the tested vehicle. Audi, Jaguar and Tesla car-makes record the lowest $\mathcal{L}_2$-gain values, as observed from Figure~\ref{fig:L2gain_Summary_Campaign_VehicleType}. In addition, manual driving is shown to be string stable with 87.5\% of the $\mathcal{L}_2$-gain estimates below the string stability threshold of $\gamma = 1$. This is mainly due to the variable spacing policy adopted by human drivers. 

\begin{figure}[!ht]
    \centering
    \includegraphics[width=0.9\textwidth]{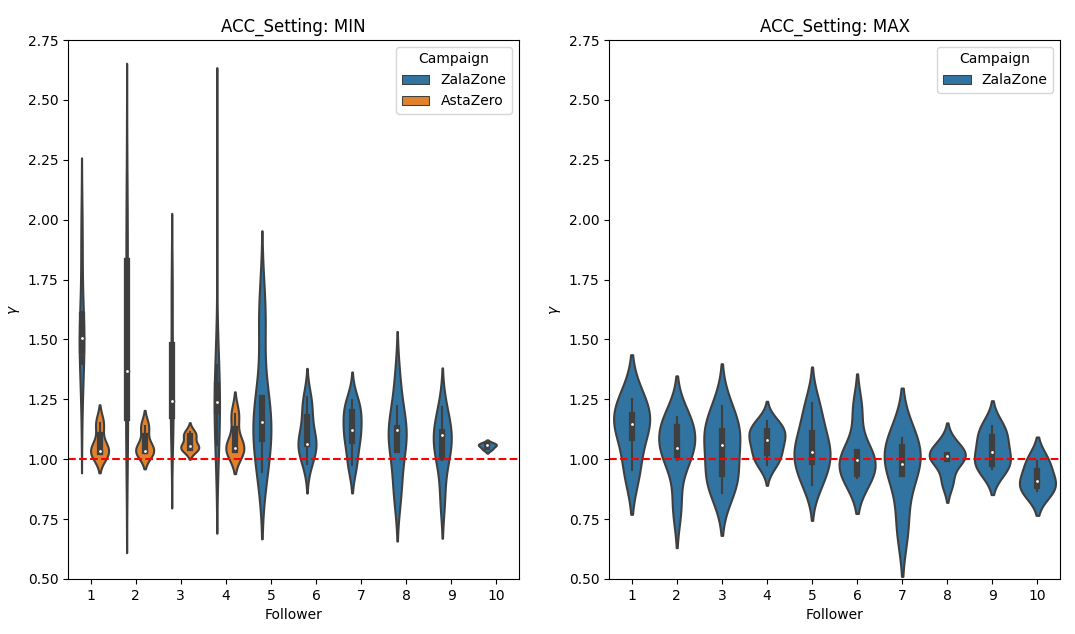}
    \caption{Distribution of estimated $\mathcal{L}_2$-gains for all leader-follower subsystems per vehicle order in the platoon for minimum (left) and maximum (right) time gap ACC settings. The red dashed line represents the threshold of string stability at $\gamma = 1$.}
    \label{fig:L2gain_Summary_Follower_Order}
\end{figure}

As demonstrated in Figure~\ref{fig:L2gain_Summary_Follower_Order}, the first vehicles inside the platoon show higher $\mathcal{L}_2$-gain estimates which show that they are more susceptible to disturbances in the driving cycle from the platoon leader and, thus, have a crucial role in dissipating such disturbances. This case is more notable for the minimum time gap ACC setting.

The findings demonstrated in Figures~\ref{fig:L2gain_Summary_Campaign_VehicleType} and~\ref{fig:L2gain_Summary_Follower_Order} second those of~\cite{Gunter2021} that commercially available ACC vehicles are string unstable. These findings are based on the assumption of linearity of the underlying system; hence, possible extensions to data-driven nonlinear analysis; specifically for variable spacing policies, are interesting for future research.

\section{Synthesis of $\mathcal{H}_\infty$-based Variable Time Gap Control Policy}
\label{SEC:ControlDesign}
In this section, a VTG control policy is designed to guarantee strict string stability during conditions away from the equilibrium profile, $s_{\text{eq}, i}$ and $v_\text{eq}$. The acceleration command and variable time gap strategy defined by our proposed controller have the following form
%The ACC system in~{\eqref{EQ:ACC_model_dyn}} defines the acceleration command of our proposed controller; however, with a variable time gap strategy of the following form %Consider the following form of the time gap command
\begin{subequations}
\begin{align}
    & a_i(t) = k_{1,i} (s_i(t) - s_0 - L_{i-1} - \tau_i(t) v_i(t)) + k_{2,i} (v_{i-1}(t) - v_i(t)) \label{eq:VTG_AccelComm}
    \\
    & \tau_i(t) = \tau^\star_i + u_i(t)
\end{align}
\end{subequations}
where $\tau^\star_i$ and $u_i(t)$ are the constant and variable time gap components respectively. It can be noted that the acceleration command in~\eqref{eq:VTG_AccelComm} has the same form as in~\eqref{EQ:ACC_model_dyn} except for the variable time-gap design.
The deviation dynamics for follower vehicle $i$ can be then derived similar to~\eqref{EQ:deviation_dyn} and are expressed as
\begin{subequations}
\begin{align}
    & \dot{\tilde{s}}_i(t) = -\tilde{v}_i(t) + \delta v_{i-1}(t) \\
    & \dot{\tilde{v}}_i(t) = k_{1,i} \tilde{s}_i(t) - (k_{1,i} \tau^\star_i + k_{2,i}) \tilde{v}_i(t) + k_{2,i} \delta v_{i-1}(t) - k_{1,i} (v_\text{eq} + \tilde{v}_i(t)) u_i(t)
    \label{EQ:velDev_nonlinearDyn}
\end{align}
\label{EQ:VTG_ACC_model_dyn}
\end{subequations}
where $\delta v_{i-1}(t)$ is considered as the disturbance from the preceding vehicle to be attenuated by the control input $u_i(t)$. The system states are denoted by the vector $\bm{x}_i(t) = \begin{bmatrix} \tilde{s}_i(t) & \tilde{v}_i(t) \end{bmatrix}^T$. Hence, the VTG component $u_i(t)$ can be designed to be reactive as a feedback control policy.

For strict input-to-state string stability, the VTG control policy for $u_i(t)$ needs to be designed to yield an $\mathcal{L}_2$-gain of $\gamma \leq 1$. 
Notably, the system is nonlinear due to the last term in~{\eqref{EQ:velDev_nonlinearDyn}}. To this end, a synthesis of nonlinear $\mathcal{H}_\infty$ controller ensues. The resulting structure considers the constant time gap component $\tau^\star_i$ as a parameter, and, in contrast, the variable time gap component $u_i(t)$ as an inner-loop control signal. This allows the proposed VTG ACC controller to adapt based on the feedback of the ego vehicle's space headway and velocity as indicators for the surrounding traffic conditions. Specifically, the variable time gap component acts in the presence of deviations away from the equilibrium profile set by $v_\text{eq}$, $s_{\text{eq}, i}$ and $\tau^\star_i$. 

The nonlinear system is represented in the following standard state-space form
\begin{subequations}
\begin{align}
    & \dot{\bm{x}} = f(\bm{x}) + g_1(\bm{x}) \bm{w} + g_2(\bm{x}) \bm{u} \\
    & \bm{z} = h(\bm{x}) + K_{12}(\bm{x}) \bm{u} 
\end{align}
\label{EQ:Hinf_ss_system}
\end{subequations}
where the subscript $i$ is dropped for brevity, $\bm{w} = \delta v_{i-1}$ denotes the exogenous disturbance, $\bm{u} = u_i$ denotes the control input, and $\bm{z} = \begin{bmatrix} \rho_s \tilde{s}_i & \rho_v \tilde{v}_i & \rho_u u_i \end{bmatrix}$ denotes the penalty variables with penalty weights of $\rho_s, \rho_v$ and $\rho_u$. Thus, the vector-valued functions are given as follows.
\begin{subequations}
\begin{align}
    & f(\bm{x}) = \begin{bmatrix}
        -\tilde{v}_i \\ k_{1,i} \tilde{s}_i - (k_{1,i} \tau^\star_i + k_{2,i}) \tilde{v}_i
    \end{bmatrix} \\
    & g_1(\bm{x}) = \begin{bmatrix}
        1 \\ k_{2, i}
    \end{bmatrix} \\
    & g_2(\bm{x}) = \begin{bmatrix}
        0 \\ -k_{1,i} (v_\text{eq} + \tilde{v}_i)
    \end{bmatrix} \\
    & h(\bm{x}) = \begin{bmatrix} \rho_s \tilde{s}_i & \rho_v \tilde{v}_i & 0 \end{bmatrix}^T \\
    & K_{12}(\bm{x}) = \begin{bmatrix} 0 & 0 & \rho_u \end{bmatrix}^T \\
    & h^T(\bm{x}) K_{12}(\bm{x}) = 0 \text{ , } K_{12}^T(\bm{x}) K_{12}(\bm{x}) = R_2 = \rho_u^2
\end{align}
\end{subequations}

The problem of designing $\bm{u}$ for string stability, as per Definition~\ref{DEF:general_string_stability}, can be formulated as an infinite-time horizon min-max optimization problem so that the $\mathcal{L}_2$-gain from $\bm{w}$ to $\bm{z}$ is $\gamma \leq 1$.
\begin{equation}
    \bm{u}^\star = \arg\min_{\bm{u}\in\mathcal{U}} \max_{\bm{w}\in\mathcal{L}_2} \frac{1}{2} \int_{0}^\infty \norm{\bm{z}(t)}^2 - \gamma^2 \norm{\bm{w}(t)}^2 dt \text{  subject to~\eqref{EQ:Hinf_ss_system}}
\label{EQ:Hinf_ocp}
\end{equation}
where $\mathcal{U} \subseteq \mathbb{R}$ is the set of admissible control inputs. Here, the objective function balances the minimization of the penalty variables, i.e.\ performance specifications, against the worst-case exogenous disturbances. Notably, the penalty variables $\bm{z}(t)$ include the deviation of the space headway away from the equilibrium profile $s_{\text{eq},i}$ and the manipulated time gap command $u_i(t)$. This motivates minimizing the space headway, as an incentive to more efficient road utilization and traffic capacity from a microscopic perspective, and the manipulated time gap command to maintain proximal operating conditions to the constant time gap setting $\tau^\star_i$.

\begin{figure}[!ht]
    \centering
    \includegraphics[width=0.75\textwidth]{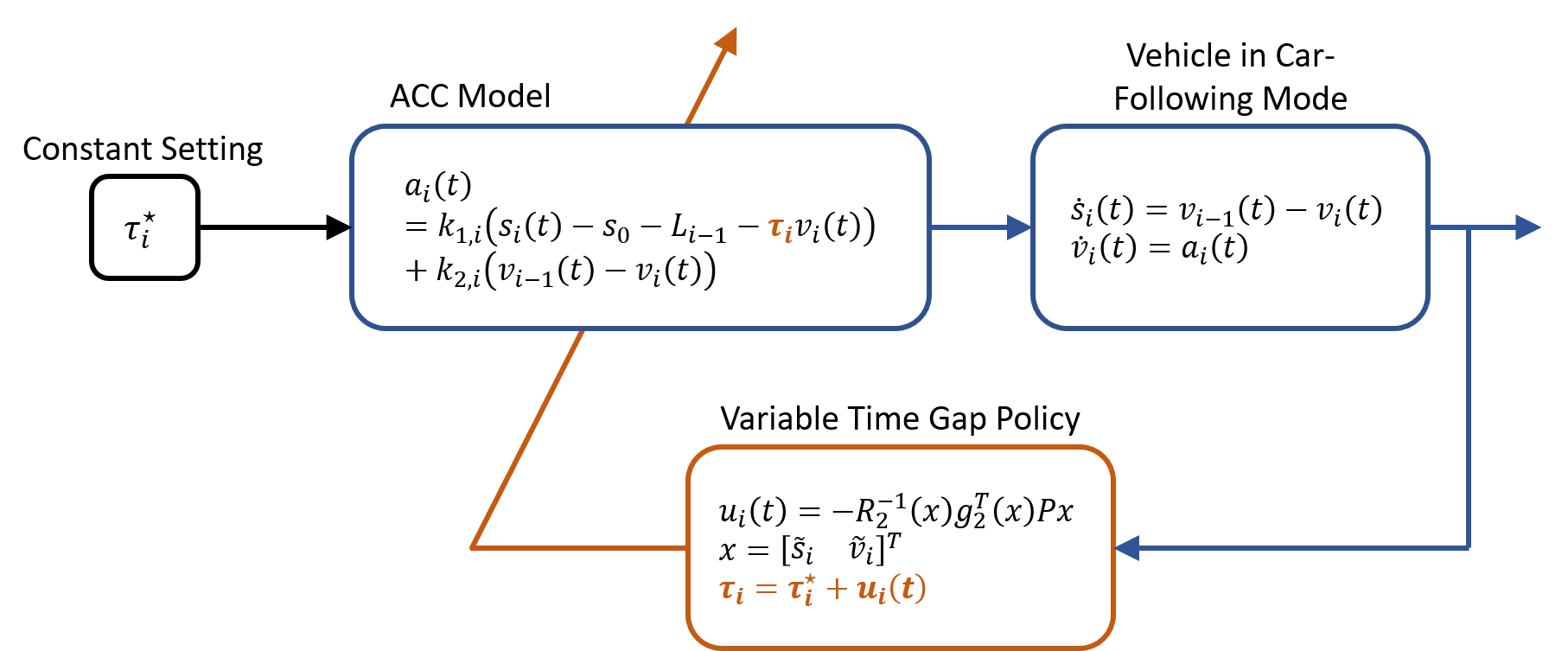}
    \caption{Block diagram of the VTG proposed feedback control policy.}
    \label{fig:VTG_block_diagram}
\end{figure}

This optimization problem is also a differential game; consisting of a two-player zero-sum game. The minimizing player controls the input $\bm{u}$; whereas the maximizing player controls the disturbance $\bm{w}$. This game has a saddle-point equilibrium solution if it has a value function $V: \mathcal{X}\mapsto\mathbb{R}$ that is positive definite and satisfies the Hamilton-Jacobi-Isaacs (HJI) equation and the optimal feedback policies of both players~\cite{Huang1995}
\begin{subequations}
\begin{align}
    & \bm{u}^\star = -R_2^{-1}(\bm{x}) g_2^T(\bm{x}) V_x \\
    & \bm{w}^\star = \frac{1}{\gamma^2} g_1^T(\bm{x}) V_x \\
    & V_x f(\bm{x}) + \frac{1}{2} h^T(\bm{x}) h(\bm{x}) + \frac{1}{2} V_x \Big( \frac{1}{\gamma^2} g_1(\bm{x}) g_1^T(\bm{x}) - g_2(\bm{x}) R_2^{-1} g_2^T(\bm{x}) \Big) = 0 
    \label{EQ:HJI}
\end{align}
\end{subequations}
where $V_x$ is the Jacobian matrix of $V(\bm{x})$. A closed-form solution for~\eqref{EQ:HJI} does not exist, thus, a numerical approximation using Taylor's series is adopted. Then, the HJI equation becomes an algebraic Ricatti equation of the following form to be solved online; since we consider the equilibrium velocity as the subsystem leader velocity $v_\text{eq} = v_{i-1}$.
\begin{equation}
    P A + A^T P + P \Bigg( \frac{1}{\gamma^2} B_1 B_1^T - B_2 R_2^{-1} B_2^T \Bigg) P + C^T C = 0 \label{EQ:Hinfty_ARE} 
\end{equation}
where $P \succeq 0$ is a symmetric positive definite matrix. The required linearized system matrices are found in \nameref{SEC:Appendix}. On another note, the algebraic Ricatti equation in~\eqref{EQ:Hinfty_ARE} is feasible for a prescribed $\gamma$ if the pair $(A, B_2)$ is stabilizable and the Hamiltonian matrix 
\begin{equation}
    H = \begin{bmatrix} 
    A & \Bigg( \frac{1}{\gamma^2} B_1 B_1^T - B_2 R_2^{-1} B_2^T \Bigg) \\ 
    -C^T C & -A^T 
    \end{bmatrix}
    \label{eq:Hinfinity_Feasibility_Condition}
\end{equation}
is dichotomic, i.e.\ has no eigenvalues on the imaginary axis. This is true for $v_\text{eq} > 0$ and for an appropiate choice of penalty weights, $\rho_s$, $\rho_v$ and $\rho_u$. The VTG ACC control strategy can be illustrated in Figure~\ref{fig:VTG_block_diagram}.
Via Figure~{\ref{fig:VTG_block_diagram}}, we can visualize the role of the constant time gap setting $\tau^\star_i$ as a parameter and the variable time gap signal $u_i(t)$ as an inner-loop control signal to guarantee string stability away from car-following equilibrium.

Furthermore, an automatic emergency braking system can incorporated as a fail-safe mechanism to guarantee collision avoidance
\begin{equation*}
    a_i(t) = a_{\text{min}}, \qquad \text{if  } \frac{v_i^2(t) -  v_{i-1}^2(t)}{2 s_i(t)} \leq a_{\text{min}}
\end{equation*}
where $a_{\text{min}} = -5 \text{m/s}^2$.

This VTG policy is robust to velocity disturbances from the leader away from the equilibrium profile; thus it yields a string-stable system in the strict sense with a worst-case $\mathcal{L}_2$-gain of $\gamma \leq 1$. Therefore, the parameter $\gamma$ prescribes the desired ``level of string stability'' in closed-loop and the smaller $\gamma$ is, the more conservative the closed-loop response is.
It also guarantees the internal stability of the considered vehicle $i$. Additionally, the constant time gap setting can be kept either at the minimum setting for benefits in traffic efficiency or at a desired setting chosen by the driver. Hence, the time gap parameter design is modularized by separating string stability from other considerations.

\section{Numerical Simulations}
\label{SEC:NumSim}
In this section, the performance of the proposed VTG ACC scheme will be investigated against the CTG ACC scheme~\cite{Milanes2014} calibrated via field platoon data. The numerical simulations are performed on real driving cycles from the OpenACC dataset to demonstrate the efficacy of the proposed VTG ACC scheme versus commercially available ACC systems. An implementation is openly available at \url{https://github.com/ivt-baug-ethz/vtg-acc}.

\subsection{Performance Indices}
In order to assess the performance of the different ACC algorithms, several indices are introduced to quantify the achieved performance. The first performance indicator is used to measure the propagation of perturbations upstream and estimate the $\mathcal{L}_2$-gain, $\gamma$, of the predecessor-follower subsystems in the worst-case scenario. It can estimated from simulated trajectories using the approach presented in Section~\ref{SEC:L2gainEst}. The estimated $\hat{\gamma}$ computed via the semi-definite program in \eqref{EQ:data_driven_L2gain_estmation} represents the string stability aspect.

Furthermore, Time-To-Collision (TTC) is used as a safety surrogate measure and is defined instantaneously as
\begin{equation}
    TTC_i(t) = \frac{s_i(t)}{v_i(t) - v_{i-1}(t)}
\end{equation}
which is valid in car-following situations where the following vehicle is faster than the preceding one, $v_i(t) > v_{i-1}(t)$. The last indicator is used to quantify the energy efficiency aspect and is the tractive energy consumption of a vehicle~\cite{Apostolakis2023}.
\begin{subequations}
\begin{align}
    & P_i(t) = \max\Big(0, 10^{-3} v_i(t) \big( F_0 + F_1 v_i(t) + F_2 v_i^2(t) + 1.03 m a_i(t) + m g \sin{\theta(t)} \big) \Big) \\
    & E_i = \frac{\int_0^T P_i(t) dt}{0.036 \int_0^T v_i(t) dt} 
\end{align}
\label{EQ:Ec}
\end{subequations}
Tractive energy consumption takes into account only tractive power demand so that it is agnostic to vehicle specifications; enabling a fair comparison between different ACC algorithms. The road is assumed horizontal; $\theta = 0$. The road-related coefficients $F_0 = 213$ N, $F_1 = 0.0861$ Ns/m, $F_2 =0.0027 \text{Ns}^2\text{/m}^2$ and the vehicle's mass $m = 1500$ kg are assumed constant over all vehicles to normalize over the different platoons and compare the performance of the utilized ACC algorithms. 

These performance indicators can also be employed for the sake of parameter tuning of the proposed VTG ACC scheme; specifically, to optimize the penalty weights $\rho_s, \rho_v, \rho_u$, and $\gamma$. Interested readers can refer to \nameref{SEC:AppendixC}.

\subsection{Driving Cycles from Field Data}
\label{SUBSEC:AstaZero}
In order to assess the performance of the proposed VTG ACC scheme, real driving cycles are used from the AstaZero campaign available in the OpenACC dataset~\cite{Makridis2021}. The field platoon data used here is acquired using minimum time gap setting for the ACC systems of all vehicles in the platoon. First, the CTG ACC model by~\cite{Milanes2014} is calibrated against each leader-follower subsystem's data trajectories. The calibration process is carried out through the following optimization program~\cite{He2022}
\begin{subequations}
\begin{align}
    & \beta^\star = \arg\min_{\beta} \gamma_a f_\text{NRMSE}(a_i) + \gamma_v f_\text{NRMSE}(v_i) + \gamma_s f_\text{NRMSE}(s_i) \\
    & \text{subject to } \beta \in \Omega_\beta \\
    & \beta = \begin{bmatrix} k_{1,i} & k_{2,i} & \tau_i \end{bmatrix}^T \\
    & f_\text{NRMSE}(y) = \frac{\sqrt{\frac{1}{T} \sum_{t=0}^T (y_\text{sim}(t) - y_\text{obs}(t))^2}}{\sqrt{\frac{1}{T} \sum_{t=0}^T y_\text{obs}^2(t)}}
\end{align}
\end{subequations}
where $f_\text{NRMSE}(y)$ is the normalized root mean squared error function used as a goodness of fit measure for the signal $y$ and $\Omega_\beta$ is the set of admissible parameters $\beta$. The simulated and observed (i.e.\ from data) trajectories of the signal $y$ are denoted by $y_\text{sim}(t)$ and $y_\text{obs}(t)$ respectively. 

It should be noted that the purpose of this calibration procedure is mainly to ensure a fair comparison between the proposed VTG ACC controller and existing ACC systems. Owing to proprietary issues, the calibrated CTG ACC model can be a valid representative of existing ACC systems.
Furthermore, the same parameter values for the calibrated CTG ACC model are used for the proposed VTG ACC controller; i.e.~gains $k_{1,i}, k_{2,i}$ and constant time gap component $\tau^\star_i$. The rationale is to isolate the proposed VTG policy as the sole influence on the ensuing results and discussions. Finally, regarding real-world application, no real-time parameter estimation for these parameters is required and the VTG ACC controller can be deployed with its standalone parameters $k_1$, $k_2$ and $\tau^\star$ as desired. 

\begin{figure}[!ht]
    \centering
    \subfloat[OpenACC Data]{
        \includegraphics[width=0.8\textwidth]{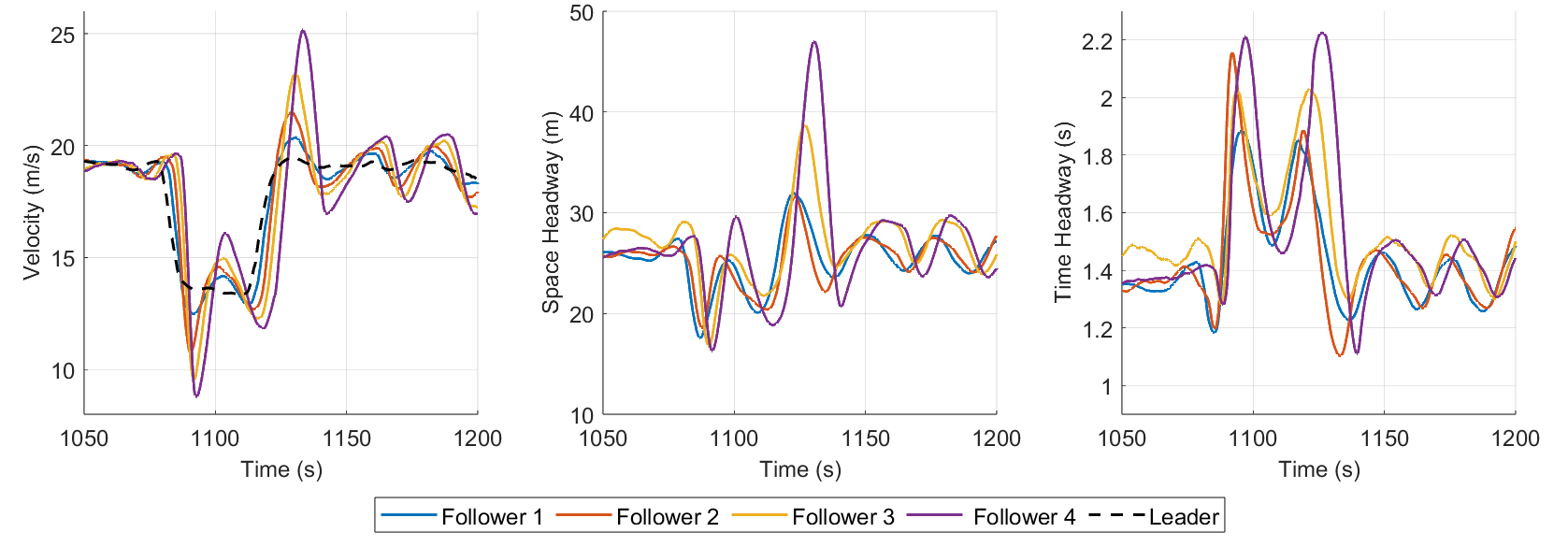}
        \label{fig:OpenACC_AstaZeroMin_Data}
    }
    \vspace{-2pt}
    \subfloat[Calibrated CTG ACC]{
        \includegraphics[width=0.8\textwidth]{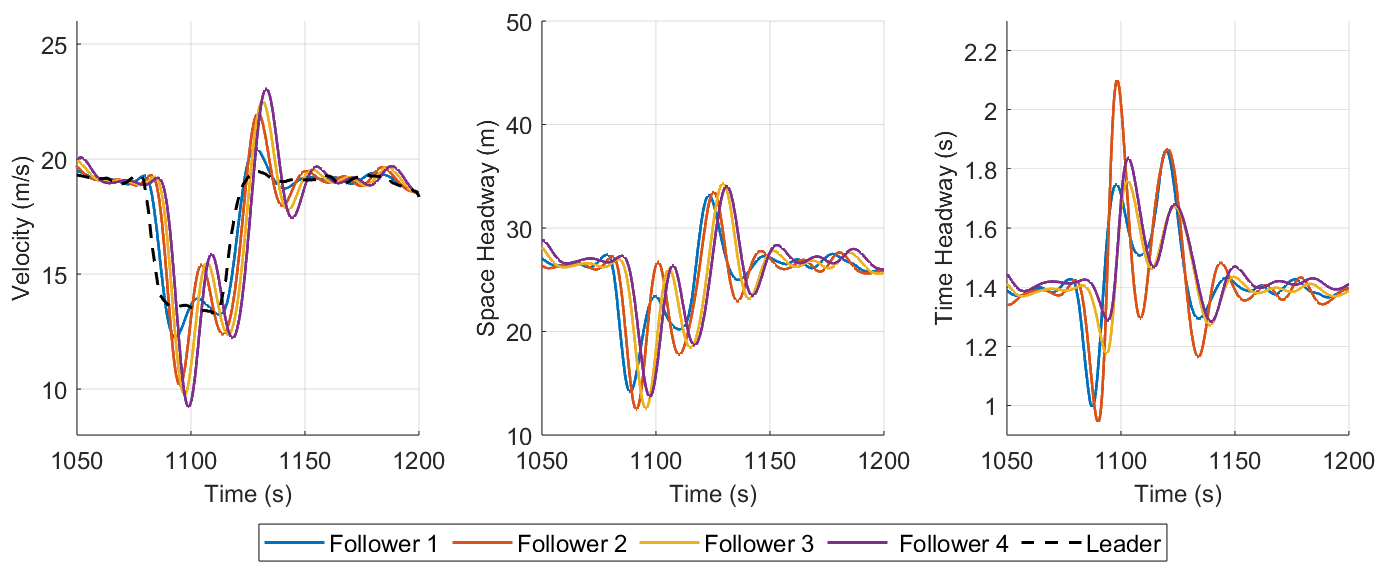}
        \label{fig:OpenACC_AstaZeroMin_ACC}
    }
    \vspace{-2pt}
    \subfloat[Proposed VTG ACC]{
        \includegraphics[width=0.8\textwidth]{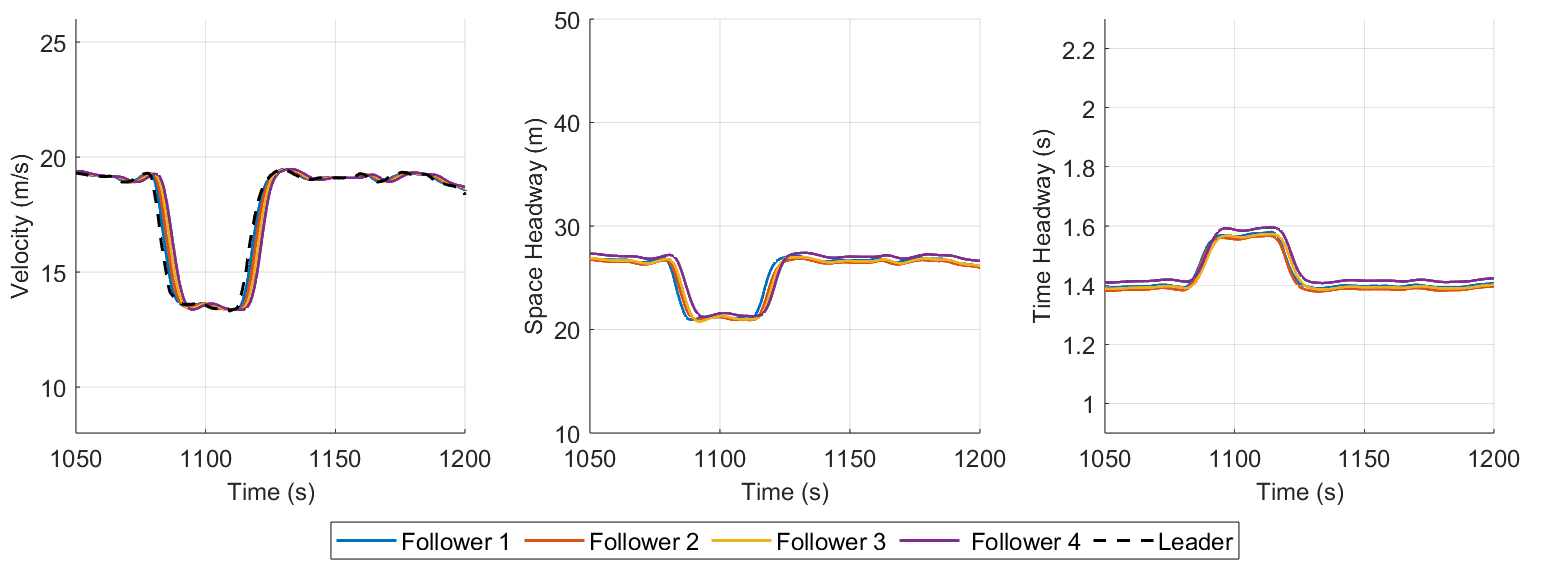}
        \label{fig:OpenACC_AstaZeroMin_VTG}
    }
    \caption{Simulated trajectories of a platoon of 5 vehicles along a straight road using the calibrated CTG ACC and the proposed VTG ACC. Platoon leader velocity is the black dashed line from the AstaZero experimental campaign. All vehicles in this platoon are using ACC with minimum time gap settings.}
    \label{fig:OpenACC_AstaZeroMin}
\end{figure}

Figure~\ref{fig:OpenACC_AstaZeroMin} shows an excerpt of the trajectories of the first and last following vehicles in the platoon. The perturbations in leader velocity are magnified in the upstream direction for the CTG ACC scheme, while the VTG ACC scheme exhibits a robust and resilient performance against these disturbances. Also, the proposed algorithm records a significant reduction in space and time headways compared to the calibrated CTG ACC and the data collected from commercially available ACC systems. Also, it is noteworthy that the proposed scheme adheres to the selected constant time headway policy $\tau^\star_i$ during stable car-following conditions and deviates away from it to attenuate any disturbances as intended.

\begin{figure}[!ht]
    \centering
    \includegraphics[width=0.85\textwidth]{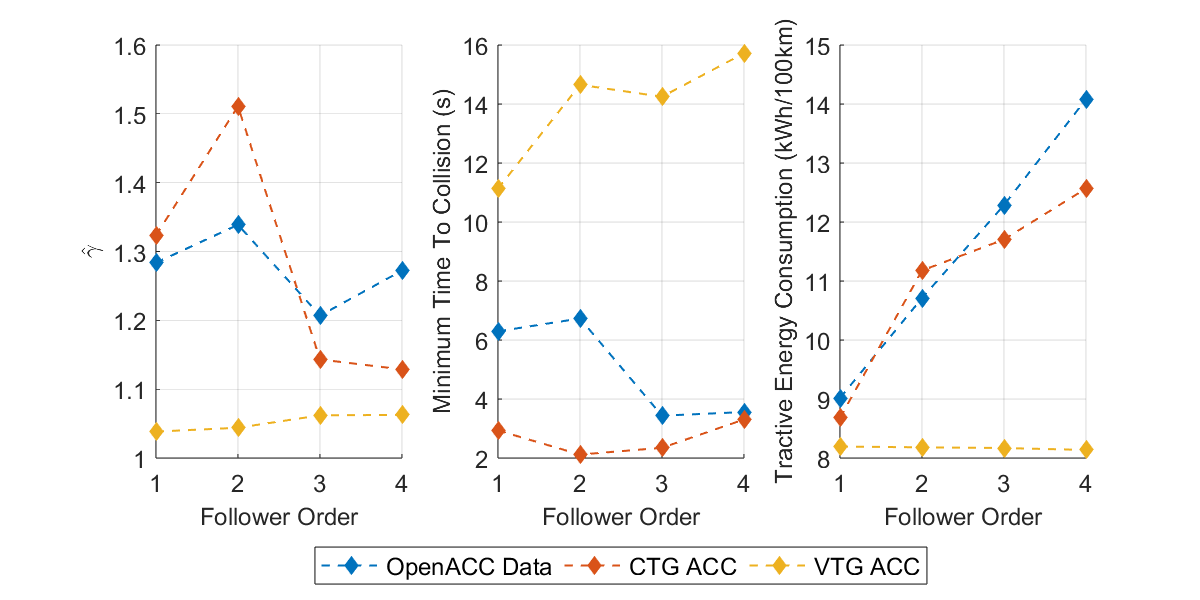}
    \caption{Performance indicators for a platoon of 5 vehicles along a straight road using the calibrated CTG ACC and the proposed VTG ACC.}
    \label{fig:OpenACC_AstaZeroMin_PerfIdx}
\end{figure}

The performance indicators for each following vehicle in the aforementioned experiment are illustrated in Figure~\ref{fig:OpenACC_AstaZeroMin_PerfIdx}. The estimated $\mathcal{L}_2$-gains for VTG ACC  are the lowest compared to the calibrated CTG ACC model and those estimated from field data; indicating better string stability. However, it should be mentioned that the estimation approach in~\eqref{EQ:data_driven_L2gain_estmation} is under the assumption of LTI systems; whereas, the VTG ACC algorithm is nonlinear. Therefore, the computed gains may be overestimated. The proposed VTG ACC algorithm achieves higher performance in terms of safety and tractive energy consumption; where the energy consumption is conformed along the entire platoon and higher TTC values are achieved. This difference is quite pronounced for the first follower since it is the immediate following vehicle to the perturbing one (i.e. platoon leader) as the proposed scheme relaxes its time gap policy to attenuate perturbations.
Further demonstrations regarding the safety impact of the proposed VTG ACC controller compared to the existing ACC systems are conducted in \nameref{SEC:AppendixB}.
In addition, the CTG ACC shows higher energy consumption overall among the platoon members due to the disturbance propagation along the platoon. Hence, safety and energy efficiency concerns raised due to string instability are relieved using the proposed approach while optimizing road space utilization by adhering to a minimum time gap during stable car-following conditions.

Furthermore, disturbances due to cut-in events are investigated in \nameref{SEC:AppendixB}.
Using the same simulation setup, the cut-in scenario is designed to evaluate the proposed controller's response to severe disturbances caused by sudden lane changes from adjacent vehicles. This maneuver results in an abrupt reduction of the gap to the directly following vehicle, generating a disturbance that propagates upstream within the platoon. The findings indicate that the proposed VTG ACC controller mitigates and dissipates the disturbance more efficiently than the CTG ACC controller, improving safety as evidenced by the TTC results and preventing the emergence of a critical situation. For a more in-depth analysis, readers are referred to \nameref{SEC:AppendixB}.

\subsubsection{Effect of Actuator Lag Dynamics and Input Delay}
Delays and lags are inherent characteristics of actuators and sensors in mechanical and control systems. Longitudinal vehicle dynamics can be modeled as a first-order lag system that defines how the vehicle powertrain responds to the acceleration command set by the ACC~\cite{Li2017, He2022}. Also, input delays are prevalent. Therefore, the low-level dynamics for vehicle $i$ can be added as follows
\begin{equation}
    \tau_{a,i} \dot{a}_i(t) + a_i(t) = a_{\text{comm}}(t-\tau_{D,i})
\end{equation}
where $\tau_{a,i}$ and $\tau_{D,i}$ represent the time lag and input time delay of the vehicle mechanical and control systems respectively. The commanded acceleration $a_{\text{comm}}$ is the computed control signal from the operating ACC system.

\begin{figure}[!ht]
    \centering
    \subfloat[Calibrated CTG ACC]{
        \includegraphics[width=0.8\textwidth]{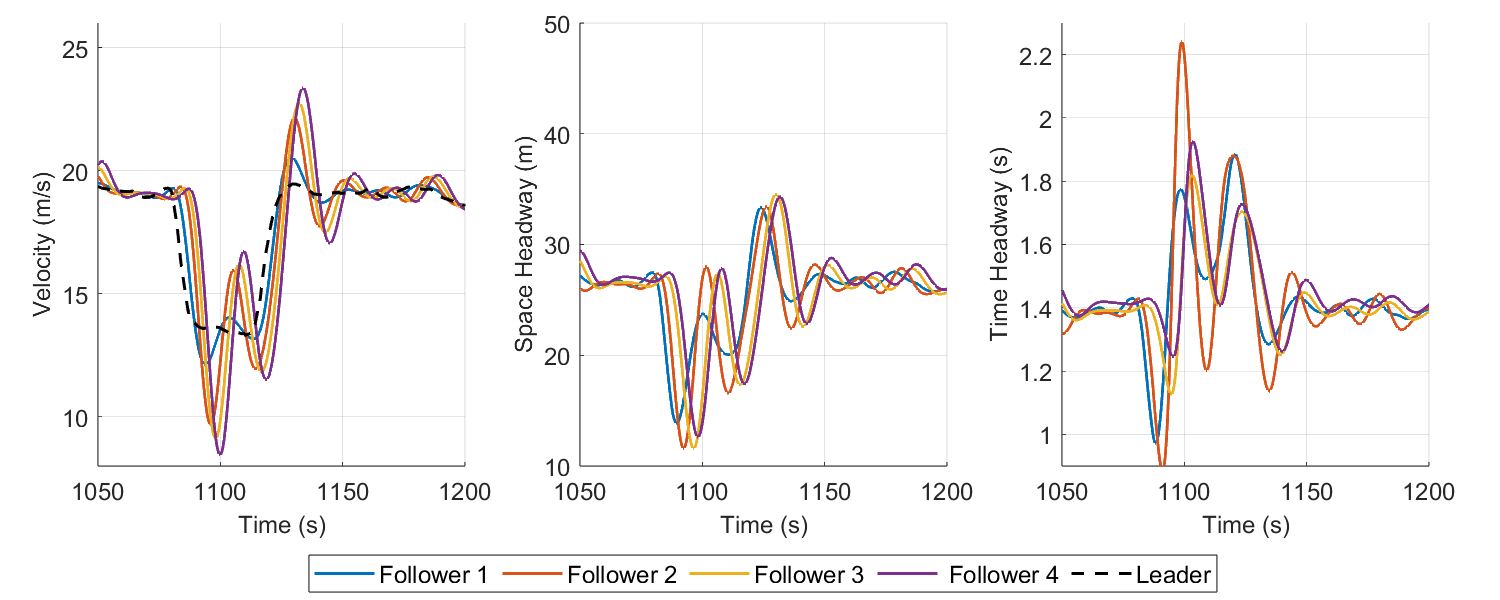}
        \label{fig:OpenACC_AstaZeroMin_ACC_Delay}
    }
    \vspace{-2pt}
    \subfloat[Proposed VTG ACC]{
        \includegraphics[width=0.8\textwidth]{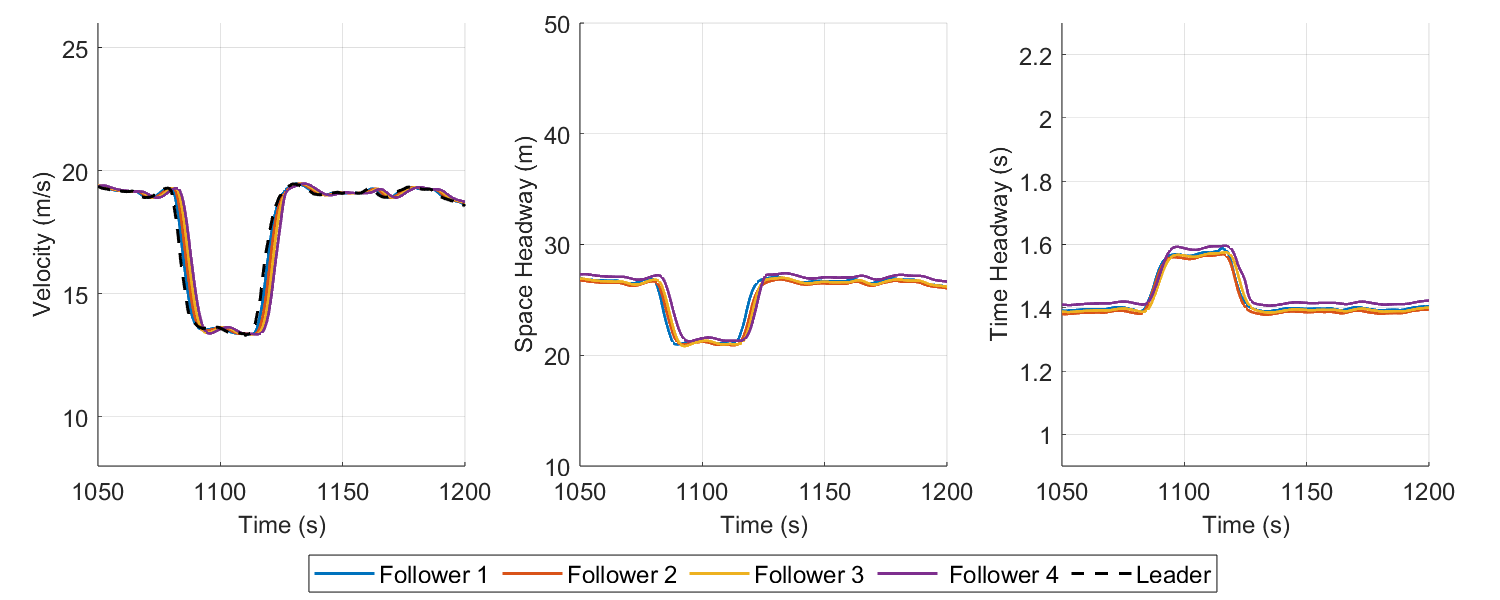}
        \label{fig:OpenACC_AstaZeroMin_VTG_Delay}
    }
    \vspace{-2pt}
    \subfloat[Performance Indicators]{
        \includegraphics[width=0.8\textwidth]{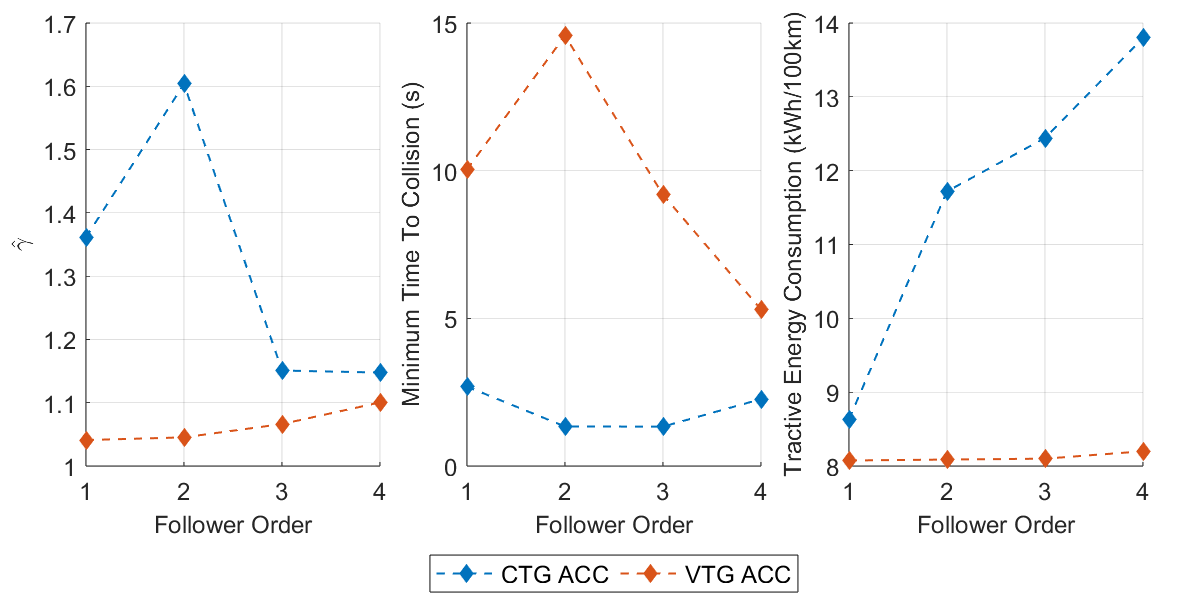}
        \label{fig:OpenACC_AstaZeroMin_PerfIdx_Delay}
    }
    \caption{Simulated trajectories of a platoon of 5 vehicles along a straight road using the calibrated CTG ACC and the proposed VTG ACC after including time lag $\tau_a = 0.1$ s and input time delay $\tau_D = 0.2$ s to simulate the effects of low-level vehicle dynamics. Platoon leader velocity is the black dashed line from the AstaZero experimental campaign. All vehicles in this platoon are using ACC with minimum time gap settings.}
    \label{fig:OpenACC_AstaZeroMin_Delay}
\end{figure}

Here, a time lag of $\tau_{a,i} = 0.1$ s and an input delay of $\tau_{D,i}=0.2$ s were used in the following simulation for all follower vehicles, i.e.\ $\forall i = 1, \dots, N$. These values were adopted based on findings from the literature after identification from real-world passenger vehicles~\cite{Ploeg2011, Ploeg2014}. The resulting vehicle trajectories, along with the performance indicators, are presented in Figure~\ref{fig:OpenACC_AstaZeroMin_Delay} for both the CTG ACC and VTG ACC schemes. The estimated $\mathcal{L}_2$-gains of the proposed scheme show little to no change relative to the no lag/delay case discussed previously. Hence, this showcases the robustness of our developed scheme in comparison to the CTG ACC scheme which shows increased $\mathcal{L}_2$-gain estimates. On the other hand, the minimum TTC values exhibit a slight decrease for the VTG ACC schemes compared to the no lag/delay case. Yet, the developed scheme shows much safer behavior than its CTG counterpart.

\begin{figure}[!ht]
    \centering
    \includegraphics[width=1.0\textwidth]{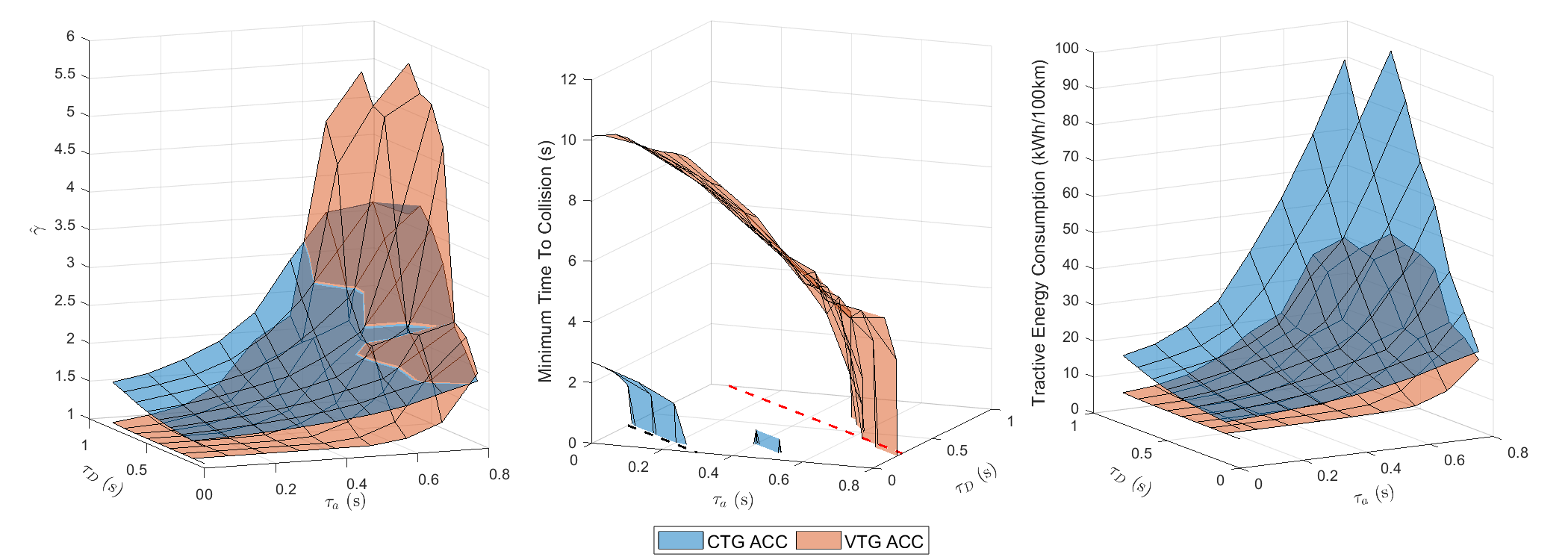}
    \caption{Average performance indicators for a platoon of 5 vehicles along a straight road using the calibrated CTG ACC and the proposed VTG ACC after varying the time lag $\tau_a$ and input time delay $\tau_D$ to simulate the effects of low-level vehicle dynamics. The black and red dashed lines indicate a total lag/delay of 0.3 and 1.05 seconds respectively.}
    \label{fig:OpenACC_AstaZeroMin_Delay_Sensitivity}
\end{figure}

Additionally, a sensitivity analysis was conducted to investigate the effect of the time lag and input time delay and evaluate their impact on system performance. The results of this analysis are shown in Figure~\ref{fig:OpenACC_AstaZeroMin_Delay_Sensitivity} after sweeping different values of time lag and input delay, up to 1 s each. For a collective delay of more than around 0.3 and 1.05 seconds (shown by the black and red dashed lines), the CTG ACC and VTG ACC algorithms yielded collisions between follower vehicles; thus an unsafe flow due to the increase of the lag/delay. Below these thresholds, the developed VTG ACC scheme shows better performance and robustness against uncertainties (i.e.\ unmodeled low-level dynamics) in terms of stability, safety, and energy efficiency, as observed in the sensitivity analysis performed. In these cases of increased lag/delay of the low-level vehicle dynamics, a compensation strategy for this delay should be added to the ACC design using predictor feedback methods as in~\cite{bekiaris2024}.

\subsection{Analysis of Different User Needs} \label{SUBSEC:UserNeeds}
In the proposed VTG strategy, there exists a constant time-gap setting $\tau^\star_i$ alongside the manipulated variable time-gap signal that was previously developed. This constant time-gap component has been previously set to a minimum setting for traffic efficiency. Another possibility might be to address user needs; where the user can select this constant time gap setting $\tau^\star_i$ as desired per their comfort level. Also, different features can be manipulated to accommodate various users as well --- such as the aggressiveness of longitudinal maneuvers through the penalty weights $\rho_s$ and $\rho_v$. We measure the aggressiveness through the rise time and settling time in response to a step change in the platoon leader's velocity. The rise time is the time taken by a follower vehicle to transition its velocity from $10\%$ to $90\%$ of the final steady-state velocity. In contrast, the settling time is the time taken to decay the transient response of the follower vehicle velocity within $\pm 1\%$ of the final steady-state velocity. Therefore, the more the rise time and/or settling time, the less aggressive the longitudinal maneuver is. Also, any difference between the rise time and settling time values could be attributed to the oscillatory response to the step change in the platoon leader's velocity; which is, in turn, another indicator for possible string instability. 
Also, the system performance was majorly influenced by the aforementioned parameters, while the effect of the gains, $k_1$ and $k_2$, was minimal. Hence, in this set of simulations, the gains, $k_1$ and $k_2$, were kept constant at $0.23$ and $0.07$. The penalty variable $\rho_u$ was kept constant at $\rho_u=1.0$ in order to satisfy the feasibility condition in \eqref{eq:Hinfinity_Feasibility_Condition}.

\begin{figure}[!ht]
    \centering
    \includegraphics[width=1.0\textwidth]{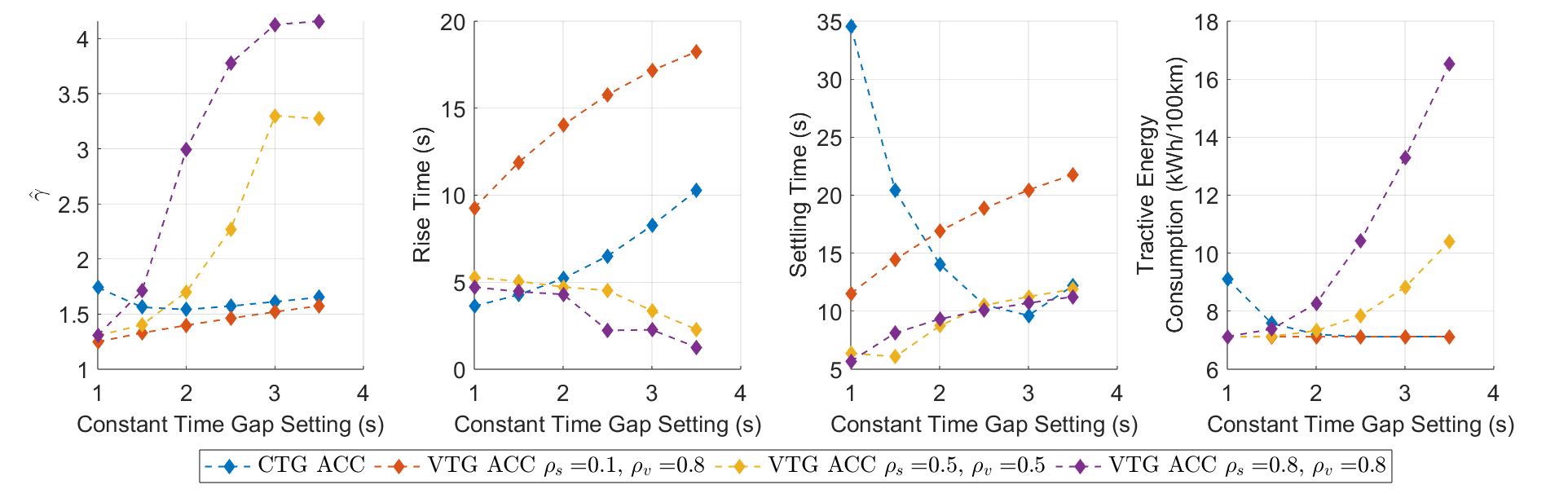}
    \caption{Average performance indicators for a platoon of 5 vehicles along a straight road using the CTG ACC and the proposed VTG ACC after varying the constant time gap setting $\tau^\star_i$ at different penalty weights, $\rho_s$ and $\rho_v$.}
    \label{fig:UserNeeds_TgVaried}
\end{figure}

Figure~\ref{fig:UserNeeds_TgVaried} shows the performance indicators as the constant time gap component $\tau^\star_i$ was varied. For penalty weights $\rho_s = 0.1$ and $\rho_v = 0.8$ (shown in red), the VTG ACC yields a lower level of aggressiveness (demonstrated by the increase in rise time and settling time) as the constant time gap setting $\tau^\star_i$ increases. This holds while relatively maintaining the same level of stability and energy efficiency. Hence, we recommend using a similar set of penalty weights for robust performance. We can also observe an upper bound for the enhanced performance of the proposed VTG ACC controller compared to the baseline CTG ACC; where both schemes achieve the same average tractive energy consumption and estimated $\mathcal{L}_2$-gains for higher constant time gap settings. This intuitively fits with previous observations that CTG ACC with long time gap settings are string stable. However, the CTG ACC scheme yields a higher level of aggressiveness through lower rise and settling times observed as $\tau^\star_i$ increases. For other penalty weights illustrated in Figure~\ref{fig:UserNeeds_TgVaried}, it can be observed that the VTG ACC controller yields an inferior performance compared to its CTG counterpart.

Another sensitivity analysis is carried out in \nameref{SEC:AppendixB} to assess the effect of the penalty weights, $\rho_s$ and $\rho_v$.

\clearpage
\section{Traffic Simulations}
\label{SEC:SUMOSim}
In this section, an investigation of the proposed scheme is carried out from a traffic flow perspective to validate our findings and the potential benefits of the proposed algorithm. Simulation of Urban Mobility (SUMO) is used for microscopic simulations. 
The proposed algorithm (VTG ACC) will be tested against the CTG ACC~\cite{Milanes2014}, two string-stable controllers from the literature~\cite{Zhou2004, Mousavi2023}, and the IDM models. Again, field platoon data from the OpenACC dataset were used to calibrate the CTG ACC model for commercially available ACC-equipped vehicles with minimum time gap settings. The same calibrated parameters are used for the VTG ACC as well. Also, manual driving data from the AstaZero campaign in the OpenACC dataset is utilized for calibration of IDM to model human driving; which will be termed ``IDM Human'' in the subsequent sections.

The approach adopted by~\cite{Zhou2004} utilizes a quadratic spacing policy inspired from human driving behavior and optimized for string stability and maximum traffic capacity. The desired space headway $s_{\text{des}, i}(t)$ according to the spacing policy can be expressed as
\begin{equation}
    s_{\text{des}, i}(t) = L_{i-1} + A + T v_i(t) + G v^2_i(t)
\end{equation}
where $A$ is the standstill distance and $T = 0.0019$ and $G = 0.0448$ are coefficients identified by the authors through offline constrained optimization. A sliding mode controller is implemented to find the follower acceleration enforcing such spacing policy. The car-following dynamics in~\eqref{EQ:general_CF_model} becomes
\begin{equation}
    f_{a,i}(s_i(t), v_i(t), v_{i-1}(t)) = \frac{\lambda}{T + 2 G v_i(t)} (s_i(t) - s_{\text{des},i}(t)) + \frac{1}{T + 2 G v_i(t)} (v_{i-1}(t) - v_i(t))
\end{equation}
where $\lambda$ is a controller gain defining the sliding surface dynamics $\dot{\epsilon} = -\lambda \epsilon$. The sliding variable $\epsilon$ is expressed through the deviation of space headway away from desired spacing $\epsilon = s_i(t) - s_{\text{des},i}(t)$. This essentially implements a VTG policy and will be referred to as ``VTG SMC''.

For the proposed approach, the penalty variables are prescribed as $\rho_s = 0.1$, $\rho_v = 2.0$, $\rho_u = 1.0$, and $\gamma = 0.95$. Furthermore, the SUMO default speed and lane-change modes were used in all the subsequent simulations.

\subsection{Ring Road Scenario}
The first traffic simulation environment is a ring road of length $L_{\text{ring}}=274$ m; where $N=10$ homogeneous vehicles are placed uniformly and initially traveling at an equilibrium flow of $v_\text{eq}=20$ m/s. This is done through a calibrated minimum time gap of $\tau^\star = 0.96770$ s.
\begin{equation}
    L_\text{ring} = N s_\text{eq} = N (L_\text{veh} + s_0 + \tau^\star v_\text{eq}) 
\end{equation}
However, to establish equilibrium flow at $20$ m/s for IDM, a length of $348$ m was needed for the ring road; where the equilibrium space headway of IDM vehicles is given as
\begin{equation}
    s_\text{eq}^\text{IDM} = \frac{s_0 + \tau_\text{IDM} v_\text{eq}}{\sqrt{1 - \Big( \frac{v_\text{eq}}{v_0} \Big)^\delta}}
\end{equation}
where $\delta$ is a model parameter.

\begin{figure}[!ht]
    \centering
    \includegraphics[width=\textwidth]{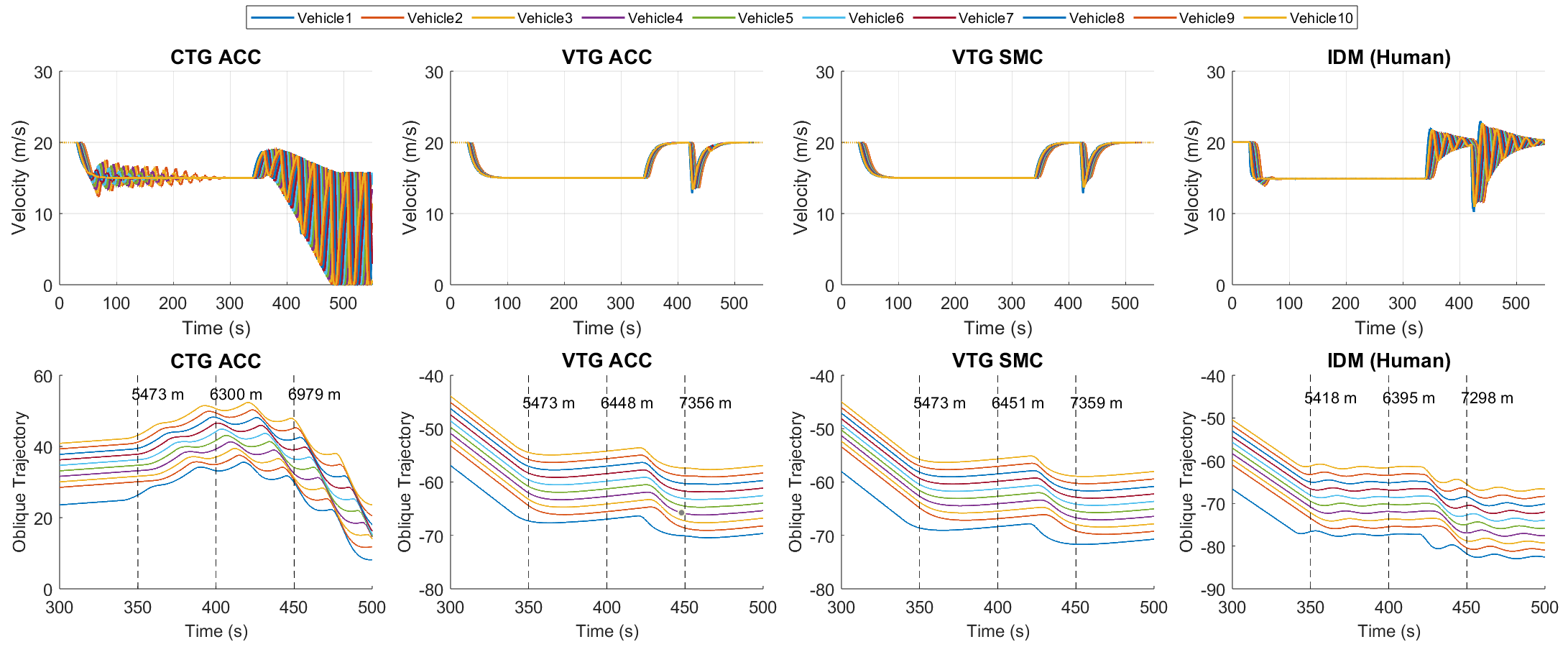}
    \caption{Velocity profiles and oblique trajectories of 10 vehicles in a ring road scenario.}
    \label{fig:Ring_Sim}
\end{figure}

\begin{figure}[!ht]
    \centering
    \includegraphics[width=0.5\textwidth]{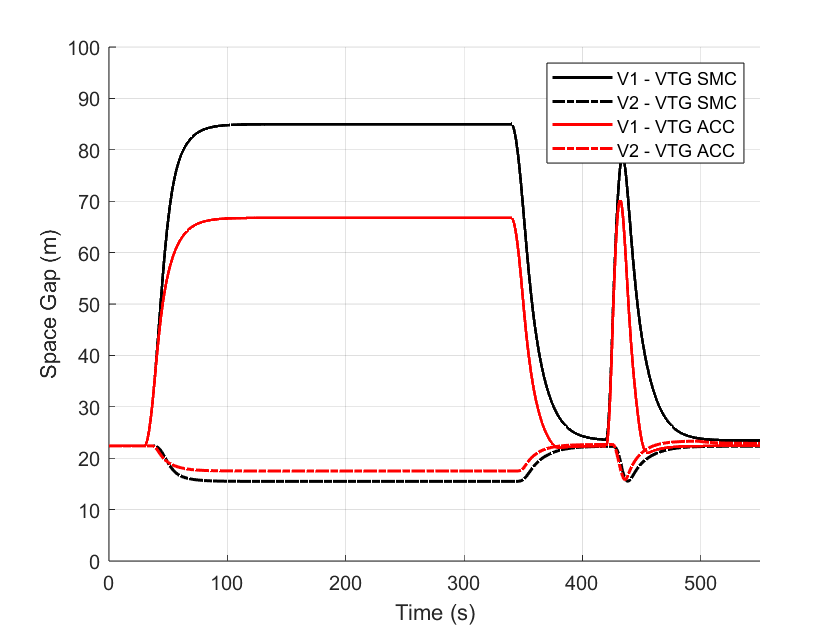}
    \caption{Spacing trajectories of 2 vehicles following the variable spacing policies of VTG SMC and VTG ACC.}
    \label{fig:Ring_Sim_Spacing}
\end{figure}

Three perturbations are introduced for Vehicle $1$ in this scenario --- deceleration to $15$ m/s at $t=30$ s, then acceleration back to $20$ m/s at $t=340$ s and, finally, sudden braking of $2 \text{m/s}^2$ at $t=420$ s. The resulting velocity trajectories and oblique time-space plots are shown in Figure~\ref{fig:Ring_Sim}. First, the string instability of the CTG ACC model is excited during the second perturbation. This induces shock waves seen in the oblique trajectories of CTG ACC vehicles in Figure~\ref{fig:Ring_Sim}; which eventually cause the vehicles to stop resulting in phantom jams or stop-and-go waves. In contrast, the VTG ACC and VTG SMC schemes show string-stable trajectories and attenuate all disturbances accordingly. Interestingly, the proposed spacing policy VTG ACC is better at achieving a balance in space gaps than the VTG SMC policy, as illustrated in Figure~\ref{fig:Ring_Sim_Spacing}. The balanced space gaps of the perturbing vehicle ``V1'' and its following vehicle ``V2'' demonstrate the potential for better road space utilization while dissipating disturbances. Regarding human driving, all three perturbations are attenuated; however, at a slower rate than the variable spacing schemes. Generally, the second and third perturbations show more aggressive effects.


\begin{figure}[!ht]
    \centering
    \subfloat[Centralized $\mathcal{H}_2$/$\mathcal{H}_\infty$ ACC scheme]{
        \includegraphics[width=0.9\textwidth]{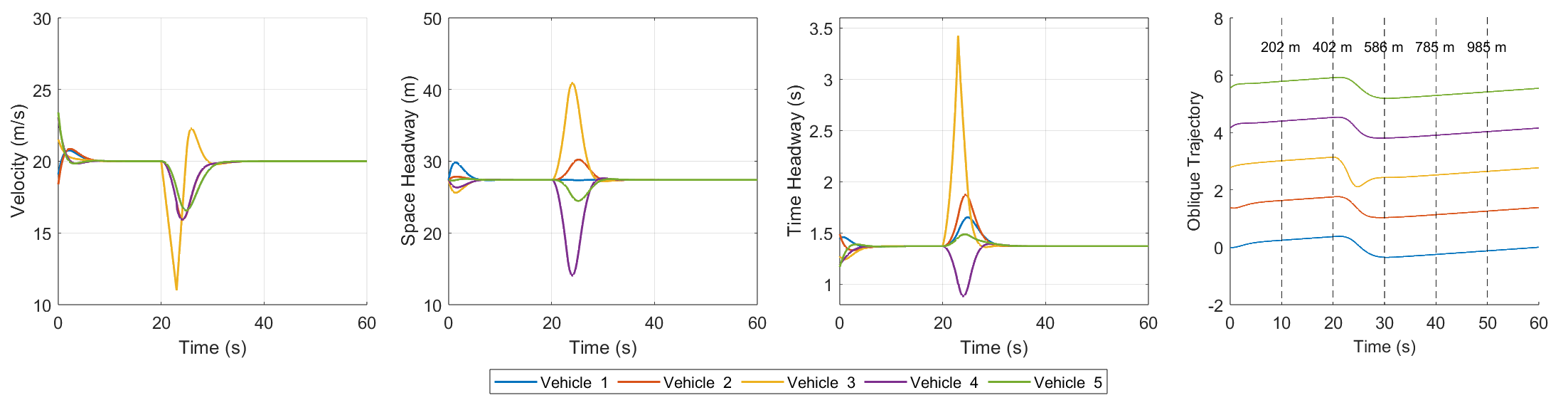}
        \label{fig:CACC_H2_Hinf_Ring}
    }
    \vspace{-2pt}
    \subfloat[Proposed VTG ACC scheme]{
        \includegraphics[width=0.9\textwidth]{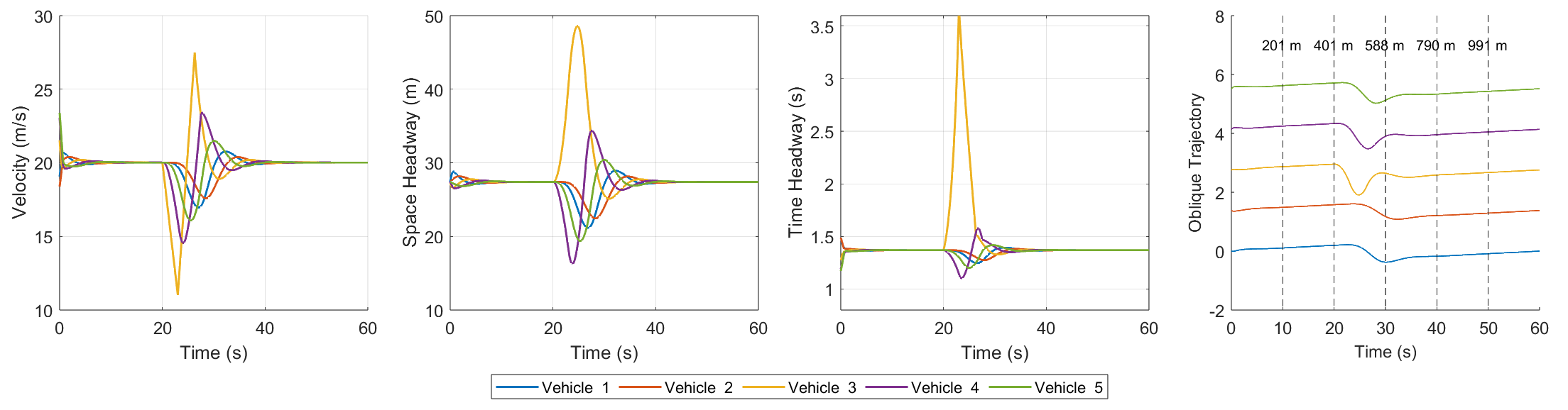}
        \label{fig:VTG_ACC_Ring}
    }
    \caption{Simulated trajectories of a platoon of 5 vehicles along a ring road.}
    \label{fig:CACC_H2_Hinf_vs_VTG_ACC}
\end{figure}

The ring road simulation scenario is repeated to investigate the performance of the proposed algorithm and the cooperative mixed $\mathcal{H}_2$/$\mathcal{H}_\infty$ algorithm introduced by~\cite{Mousavi2023}. Here, we consider homogeneous platoons of ACC-enabled vehicles for the implementation of the centralized controller by~\cite{Mousavi2023} for comparison. This test is carried out for a platoon of $N=5$ vehicles in a ring road of length $L_{\text{ring}}=137$ m as well as the calibrated minimum time gap of $\tau^\star=0.96770$ s. Initial perturbations in the vehicles' velocities are introduced as well as a sharp deceleration of $-3 \text{ m/s}^2$ lasting $3$ s for one random vehicle --- 'Vehicle 3' in Figure~\ref{fig:CACC_H2_Hinf_vs_VTG_ACC}.

As illustrated in Figure~\ref{fig:CACC_H2_Hinf_vs_VTG_ACC}, both schemes can dissipate perturbations as they are string-stable by design. However, the centralized ACC algorithm shows lower changes in space and time headway relative to the proposed VTG ACC algorithm. The reason is that the former scheme assumes a centralized cooperative structure; therefore, has global information about all platoon vehicles. In contrast, our proposed control scheme is completely decentralized via predecessor-follower topology and only needs sensor measurements of spacing and lead velocity; thus, it is more computationally efficient with comparable performance.  Interestingly, with the heightened interest in connected vehicles, our proposed scheme can be extended for connected vehicles which is left for future research.

\subsection{Highway Merging Scenario}
In this scenario, a single-lane highway of length $1$ km has an on-ramp merging at position $600$ m. The maximum flow rate of vehicles at the mainline is $2000$ veh/h while the on-ramp creates a demand of $100$ veh/hr. The vehicles' arrival on both the highway and the on-ramp are modeled as independent Poisson processes with the corresponding flow rate. The vehicles on the mainline have a steady-state speed of $30$ m/s. The aggressive joining of vehicles from the on-ramp to the mainline may cause shock waves downgrading the system's performance. The default lane change model of SUMO is used in this simulation scenario.

In this set of simulations, another benchmark ACC controller by~\cite{Wang2004} is brought into the comparison; where the desired spacing policy is based on the Greenshields fundamental diagram, and the acceleration command is designed via the sliding mode technique. This benchmark controller is a traffic-centric approach and will be referred to as ``FD-VTG SMC''. The desired spacing policy is expressed as follows.

\begin{figure}[!ht]
    \centering
    \includegraphics[width=0.9\textwidth]{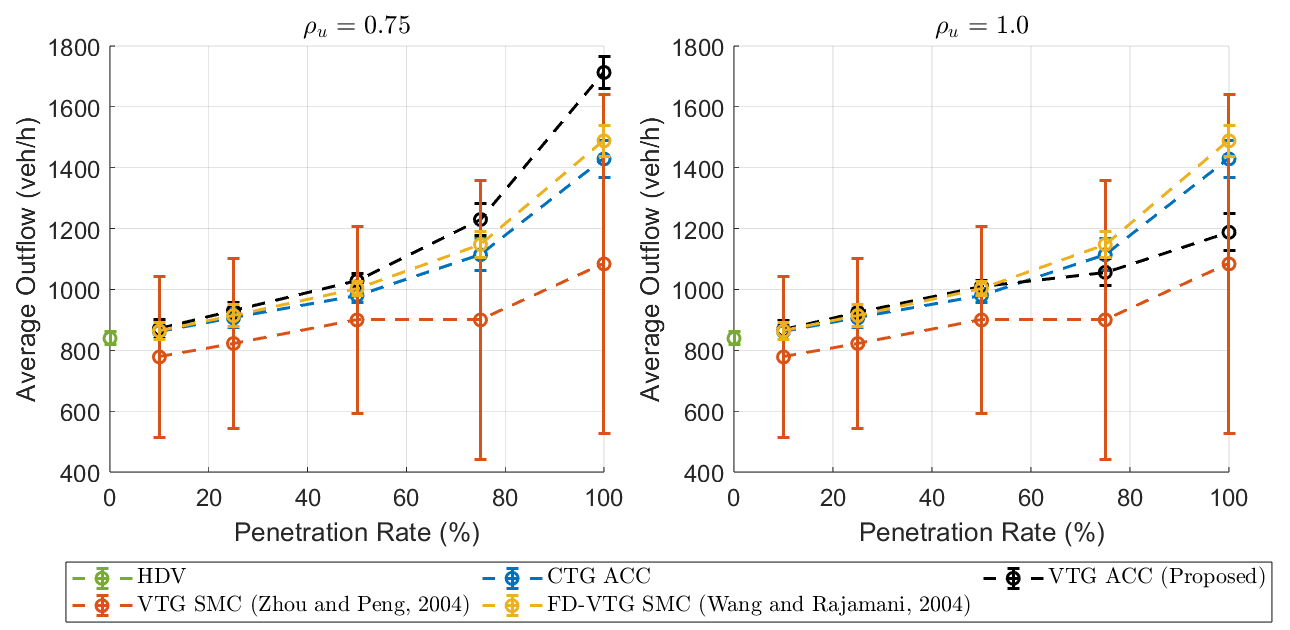}
    \caption{Average outflow from the highway merge network for different penetration rates of ACC-equipped vehicles. Error bars represent the standard deviation of the outflow from the network across the 10 runs.}
    \label{fig:Merge_Sim_Mixed_Pr_Outflow}
\end{figure}

\begin{figure}[!ht]
    \centering
    \includegraphics[width=0.5\textwidth]{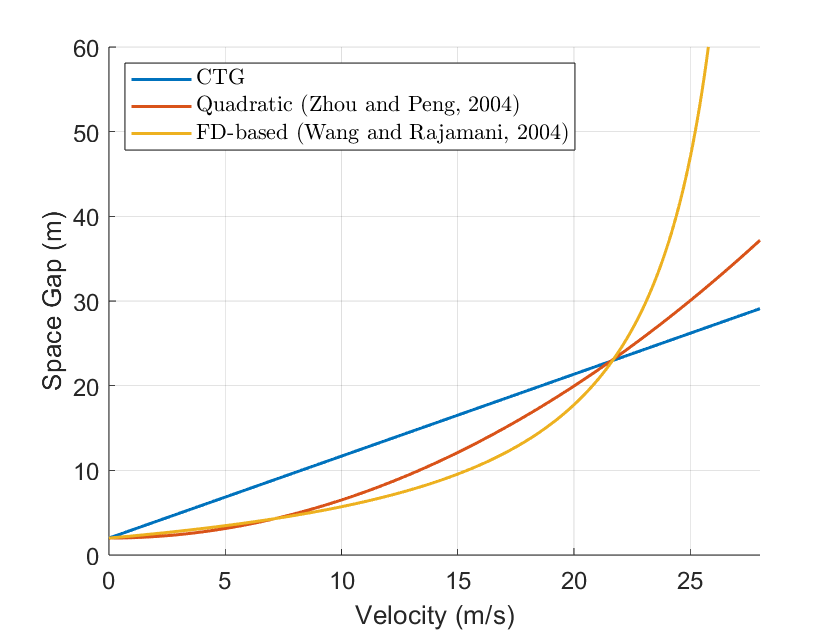}
    \caption{Constant time gap (CTG), quadratic, and fundamental-diagram-based spacing policies during stable car-following conditions.}
    \label{fig:spacing_policy_ACC_vs_SMC}
\end{figure}

A mixed traffic flow is created by varying the penetration rate, $p$, of the ACC-equipped vehicles; i.e.\ controlled by either CTG ACC, VTG ACC, VTG SMC, or FD-VTG SMC. The calibrated CTG ACC algorithm represents the commercially-available ACC systems and the calibrated IDM algorithm represents manually driven vehicles. Here, the performance of the traffic system is evaluated by the average outflow from the network versus penetration rates of $p = [0, 10\%, 25\%, 50\%, 75\%, 100\%]$; which is summarized by Figure~\ref{fig:Merge_Sim_Mixed_Pr_Outflow}. These are obtained by averaging $10$ $600$-second runs.
The proposed VTG ACC scheme for $\rho_u = 0.75$ yields an improved traffic outflow in comparison to CTG ACC, VTG SMC and FD-VTG SMC algorithms. Interestingly, both VTG SMC and FD-VTG SMC algorithms achieve lower outflow due to larger space gaps required by the adopted quadratic spacing policy for speeds higher than around $21$ m/s; as shown in Figure~\ref{fig:spacing_policy_ACC_vs_SMC}. Thus, the improvement in the traffic outflow exhibited by the proposed algorithm can be attributed to (a) dissipating perturbations created by the merging bottleneck and (b) adhering to a constant minimum time gap policy during stable car-following conditions, which shows better road space utilization and traffic efficiency.
%The proposed VTG ACC scheme for $\rho_u = 0.75$ yields an improved traffic outflow in comparison to both the CTG ACC and the VTG SMC algorithms. Interestingly, the VTG SMC achieves lower outflow due to larger space gaps required by the adopted quadratic spacing policy for speeds higher than around $21$ m/s; as shown in Figure~\ref{fig:spacing_policy_ACC_vs_SMC}. Thus, the improvement in the traffic outflow exhibited by the proposed algorithm can be attributed to (a) dissipating perturbations created by the merging bottleneck and (b) adhering to a constant minimum time gap policy during stable car-following conditions which shows better road space utilization and traffic efficiency.

Notably, the proposed VTG ACC algorithm yields lower standard deviations (represented by the error bars in Figure~\ref{fig:Merge_Sim_Mixed_Pr_Outflow}) than the VTG SMC policy; suggesting a more consistent control policy in various traffic conditions. It is important to note that both the quadratic and fundamental-diagram-based spacing policies, underlying the VTG SMC and the FD-VTG SMC algorithms, respectively, are not commercially implemented on today's roads. Such policies dictate shorter space gaps at low speeds, as observed in Figure~\ref{fig:spacing_policy_ACC_vs_SMC}, which might improve efficiency at the expense of possible safety during congested conditions. Existing ACC systems follow a more relaxed policy than the CTG policy at low speeds~\cite{Xiao2017}; which may be attributed to compensating possible perception inaccuracies. Additionally, the proposed VTG ACC regulates the commanded spacing and time gap (in contrast to the other ACC algorithms) through the feedback control design; hence, it is reactive to traffic conditions as measured by the states of the ego vehicle. It should be pointed out that these results may be affected by the nature of the SUMO lane-changing model since it assumes that lane change occurs instantaneously after the decision is taken. This aspect is intrinsically different from real-world operations. Another limitation might be the higher level of heterogeneity between human drivers on real-life roads compared to the conducted microscopic simulations. In our simulations, different desired speeds along with driver imperfection factors were implemented for the different human-driven vehicles in order to simulate some heterogeneity between human drivers.

\begin{figure}[!ht]
    \centering
    \subfloat[Time-Space Diagrams]{
    \includegraphics[width=0.95\textwidth]{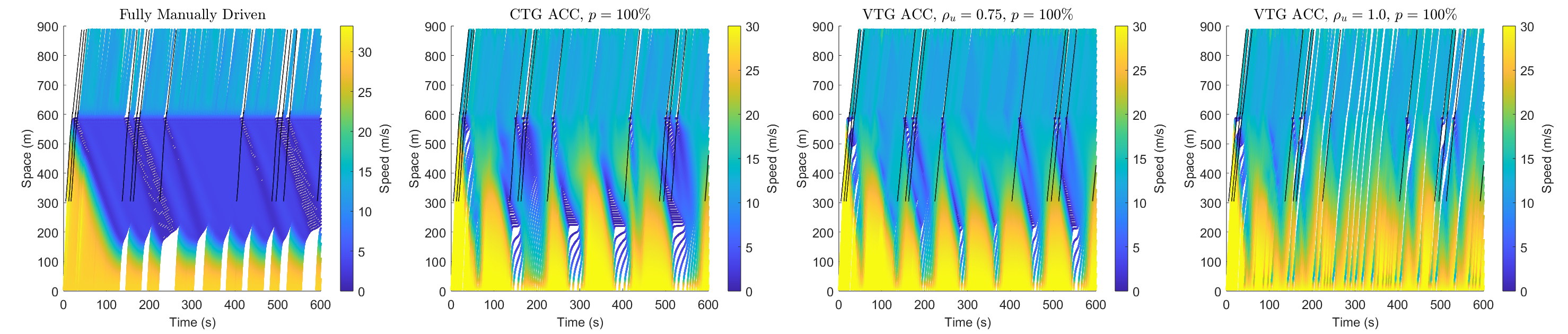}
    }
    \vspace{-2pt}
    \subfloat[Traffic Response]{
    \includegraphics[width=0.95\textwidth]{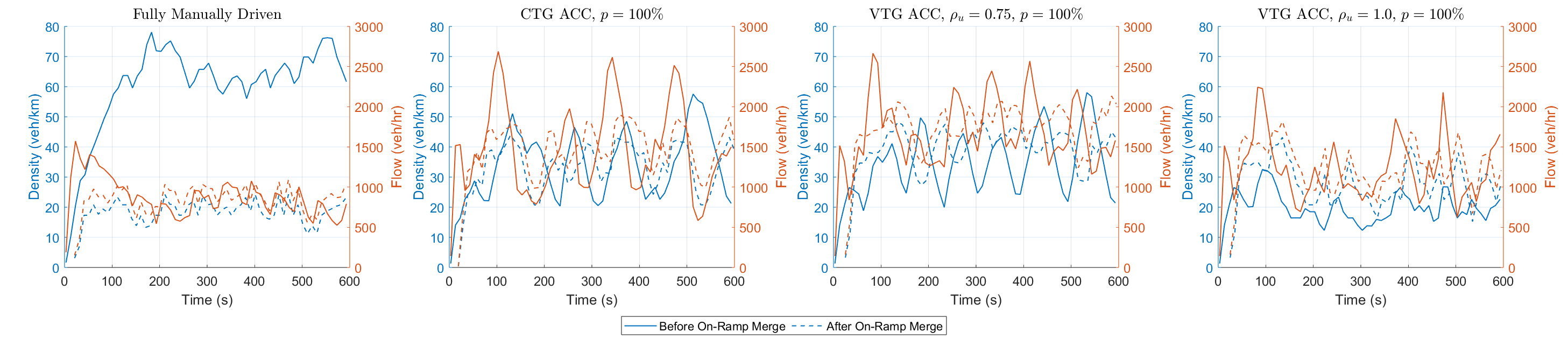}
    }
    \caption{Time-space diagrams and traffic response in a highway merging network using the calibrated CTG ACC and the proposed VTG ACC.}
    \label{fig:Merge_Sim_TSD_Traffic}
\end{figure}

On the other hand, for the case of $\rho_u = 1.0$, the VTG ACC algorithm is more conservative with respect to string stability enforcement and minimizing the manipulated time-gap component $u_i(t)$ compared to the case of $\rho_u = 0.75$. This can be further observed in Figure~\ref{fig:Merge_Sim_TSD_Traffic} where the time-space diagrams are shown for $p = 100\%$ and fully manually driven traffic flow. The VTG ACC algorithm with $\rho_u = 1.0$ shows fewer perturbations and shock-waves formed compared to $\rho_u = 0.75$. However, the traffic volume yielded after the on-ramp merging bottleneck is higher for $\rho_u = 0.75$ due to a more relaxed VTG policy.

Additionally, the performance indicators for safety and energy consumption are computed for each vehicle in the network and the ratio of improvement with respect to the calibrated CTG ACC averaged over all vehicles is computed. As illustrated in Table~\ref{TAB:Merge_Sim_Improvement_PerfIdx}, the maximum improvement in minimum TTC per vehicle is observed for the case of $\rho_u = 1.0$ as string stability is conservatively enforced; whereas more improvement is seen in tractive energy consumption for $\rho_u = 0.75$. This can be attributed to the more relaxed VTG policy which gives rise to a more smooth accommodation of the flow of merging vehicles with relaxed time headways. Hence, an adjustment of the penalty variable $\rho_u$ may result in different behaviors since it allows various degrees of relaxation of the CTG policy and string stability enforcement.

\begin{table}[!ht]
\centering
\caption{Average percentage of improvement of performance indicators for the proposed VTG ACC scheme in comparison to the calibrated CTG ACC scheme averaged per vehicle and on the 10 runs.}
\label{TAB:Merge_Sim_Improvement_PerfIdx}
\begin{tabular}{l|cc|cc}
\multirow{2}{*}{Penetration Rate} & \multicolumn{2}{c|}{VTG ACC, $\rho_u = 0.75$} & \multicolumn{2}{c}{VTG ACC, $\rho_u = 1.0$} \\ %\cline{2-5} 
                                  & minimum TTC             & $E_c$               & minimum TTC           & $E_c$                \\ \hline
$10\%$                            & $7.16\%$                & $-3.56\%$           & $48.18\%$             & $4.21\%$             \\
$25\%$                            & $9.83\%$                & $10.01\%$           & $56.03\%$             & $3.67\%$             \\
$50\%$                            & $14.19\%$               & $74.81\%$           & $55.04\%$             & $6.01\%$             \\
$75\%$                            & $15.46\%$               & $252.63\%$          & $91.66\%$             & $0.35\%$             \\
$100\%$                           & $62.05\%$               & $187.07\%$          & $261.35\%$            & $-9.49\%$            \\ \hline
\end{tabular}
\end{table}

% \begin{tabular}{l|cc|cc}
% \multirow{2}{*}{Penetration Rate} & \multicolumn{2}{c|}{VTG ACC, $\rho_u = 0.75$} & \multicolumn{2}{c}{VTG ACC, $\rho_u = 1.0$} \\ %\cline{2-5} 
%                                   & minimum TTC             & $E_c$               & minimum TTC           & $E_c$               \\ \hline
% $10\%$                            & $2.1\%$                 & $2.5\%$             & $10\%$                & $1.4\%$             \\
% $25\%$                            & $4.4\%$                 & $42\%$              & $25\%$                & $-1.3\%$            \\
% $50\%$                            & $5.7\%$                 & $161\%$             & $42\%$                & $-10\%$             \\
% $75\%$                            & $9.4\%$                 & $533\%$             & $105\%$                & $-0.69\%$           \\
% $100\%$                           & $26\%$                  & $297\%$             & $259\%$                & $-13\%$             \\ \hline
% end{tabular}
\section{Conclusion}
\label{SEC:Conc}

In this article, a variable time gap strategy is designed to guarantee string stability in the strict sense for vehicle platoons. The control architecture is modular; where a constant time gap component $\tau^\star$ and a variable time gap component $u(t)$ are designed as the control policy for disturbance attenuation. Strict string stability is guaranteed through nonlinear $\mathcal{H}_\infty$ control to dissipate perturbations from the leading vehicle to the prescribed penalty variables; space headway and ego vehicle's velocity. The string stability problem is decoupled from equilibrium flow dictated by $\tau^\star$; which may define driver comfort settings or a minimum time gap value for traffic efficiency. The performance of the proposed scheme is validated through numerical simulations compared to calibrated models representing commercially available ACC systems and manually driven vehicles; available in the OpenACC dataset, as well as variable spacing policies from the literature. Also, traffic simulations are carried out to demonstrate the efficacy of each model in both a ring road scenario and a highway merging scenario. Notably, under a penetration rate of $10\%$, the proposed VTG ACC policy improved traffic outflow by $1.05\%$ on average compared to the CTG ACC policy (representing commercially available ACC systems); as well as relieving safety and energy efficiency concerns with an improvement of $2.1\%$ and $2.5\%$ respectively.

\section*{Acknowledgments}
 The authors would like to thank Dr.~Yifan Zhang for the insightful discussions on car-following models.
\section*{Appendix A} \label{SEC:Appendix}
The deviation dynamics in \eqref{EQ:deviation_dyn} are derived as follows. 
\begin{subequations}
\begin{align}
    \dot{\tilde{s}}_i &= v_{i-1} - v_i \\
    & = (v_{i-1} - v_\text{eq}) - (v_i - v_\text{eq}) \\
    & = -\tilde{v}_i + \delta v_{i-1}
\end{align}
\label{EQ:spacing_deviation_dyn}
\end{subequations}
\begin{subequations}
\begin{align}
    \dot{\tilde{v}}_i &= k_{1,i} (s_i(t) - s_0 - L_{i-1} - \tau_i v_i(t)) + k_{2,i} (v_{i-1}(t) - v_i(t)) \\
    & =  k_{1,i} (s_i(t) - s_0 - L_{i-1} - \tau_i (\tilde{v}_i + v_\text{eq}) ) + k_{2,i} (-\tilde{v}_i + \delta v_{i-1}) \\
    & = k_{1,i} (s_i(t) - s_0 - L_{i-1} - \tau_i v_\text{eq}) - k_{1,i} \tau_i \tilde{v}_i - k_{2,i} \tilde{v}_i + k_{2,i} \delta v_{i-1} \\
    & = k_{1,i} \tilde{s}_i - k_{1,i} \tau_i \tilde{v}_i - k_{2,i} \tilde{v}_i + k_{2,i} \delta v_{i-1}
\end{align}
\label{EQ:velocity_deviation_dyn}
\end{subequations}

The linearized system matrices required for the algebraic Riccati equation in \eqref{EQ:Hinfty_ARE} can be described as follows.
\begin{subequations}
\begin{align}
    & f(\bm{x}_i) = A \bm{x}_i + \hat{f}^{[2+]}(\bm{x}_i) \text{ ; } A = \begin{bmatrix} 0 & -1 \\ k_1 & -(k_1 T_g^\star + k_2) \end{bmatrix} \\
    & g_1(\bm{x}_i) = B_1 + \hat{g}_1^{[1+]}(\bm{x}_i) \text{ ; } B_1 = \begin{bmatrix} 1 \\ k_2 \end{bmatrix} \\
    & g_2(\bm{x}_i) = B_2 + \hat{g}_2^{[1+]}(\bm{x}_i) \text{ ; } B_2 = \begin{bmatrix} 0 \\ -k_1 v_e \end{bmatrix} \label{EQ:onlyAssump1}\\
    & h(\bm{x}_i) = C \bm{x}_i + \hat{h}^{[2+]}(\bm{x}_i) \text{ ; } C = \begin{bmatrix} \rho_s & 0 \\ 0 & \rho_v \\ 0 & 0 \end{bmatrix} \\
    & V(\bm{x}_i) = \bm{x}_i^T P \bm{x}_i + \hat{V}^{[3+]}(\bm{x}_i) \label{EQ:onlyAssump2}
\end{align}
\end{subequations}
where the $n^{\text{th}}$-order and higher-order terms of the function $(\cdot)(\bm{x}_i)$ are denoted by $\hat{(\cdot)}^{[n+]}(\bm{x}_i)$ and $P \succeq 0$ is a symmetric positive definite matrix. It should be noted that, in our setting, the only assumptions made are~\eqref{EQ:onlyAssump1} and~\eqref{EQ:onlyAssump2}, and the rest are exact expressions.

%%%%%%%%%%%%%%%%%%%%%%%%%%%%%%%%%%%%%%%%%%%%%%%%%%%%%%%%%%%%%%%%%%%%%%%%%%%%%%%%%%%%%%%%%%%
%%%%%%%%%%%%%%%%%%%%%%%%%%%%%%%%%%%%%%%%%%%%%%%%%%%%%%%%%%%%%%%%%%%%%%%%%%%%%%%%%%%%%%%%%%%
\section*{Appendix B} \label{SEC:AppendixB}
\subsection*{Safety Impact: Numerical Simulations using Field Platoon Data}
\begin{figure}[!ht]
    \centering
    \includegraphics[width=0.9\textwidth]{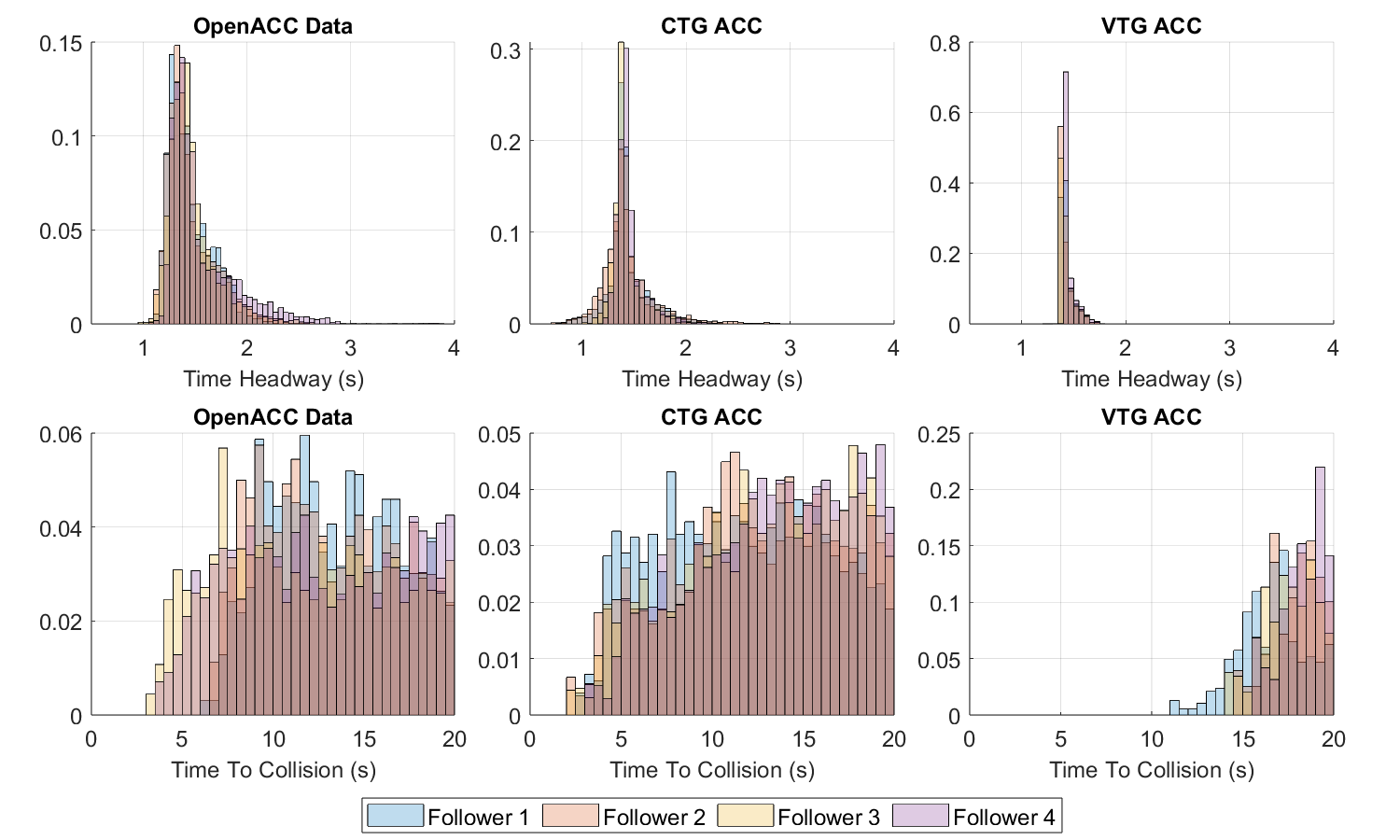}
    \caption{Distribution of time headway and time to collision for followers in a platoon of 5 vehicles along a straight road. Platoon leader is simulated using leader trajectory from the AstaZero experimental campaign. All vehicles in this platoon are using ACC with minimum time gap settings.}
    \label{fig:AstaZero_TTC_TimeHeadway_Distribution}
\end{figure}

First, an assessment of the safety impact of the developed VTG ACC controller is carried out. In Figure~\ref{fig:AstaZero_TTC_TimeHeadway_Distribution}, we show the distribution of the minimum TTC values for the proposed VTG ACC scheme in comparison to existing ACC systems from the AstaZero campaign in the OpenACC dataset (with the same setup as Section~\ref{SUBSEC:AstaZero}). Using the proposed VTG ACC algorithm, the time headway is bounded from below by the desired time gap setting; thus, enhancing its safety as shown by the distribution of the TTC values. In addition, the headway values from commercially available ACC systems and calibrated CTG ACC scheme show high variance; thus, it would be ``less safe'' than our proposed scheme as evidenced by the TTC values as a surrogate safety measure.

\begin{table}[!ht]
\centering
\caption{Surrogate safety measures for followers in a platoon of 5 vehicles along a straight road. Platoon leader is simulated using leader trajectory from the AstaZero experimental campaign. All vehicles in this platoon are using ACC with minimum time gap settings.}
\label{TAB:SSM_OpenACC_AstaZero}
\begin{tabular}{l|c|c|c}
                & \begin{tabular}[c]{@{}c@{}}Minimum TTC\\ (s)\end{tabular} & \begin{tabular}[c]{@{}c@{}}Time-Exposed TTC\\ (s)\end{tabular} & \begin{tabular}[c]{@{}c@{}}Maximum DRAC\\ ($\text{m/s}^2$)\end{tabular} \\ \hline
OpenACC Dataset & 3.43                                                      & 4.5                                                            & 0.45                                                                    \\
CTG ACC         & 2.12                                                      & 22.7                                                           & 0.54                                                                    \\
VTG ACC         & 11.13                                                     & 0                                                              & 0.08                                                                    \\ \hline
\end{tabular}
\end{table}

\begin{figure}[!ht]
    \centering
    \subfloat[OpenACC Data]{
        \includegraphics[width=0.8\textwidth]{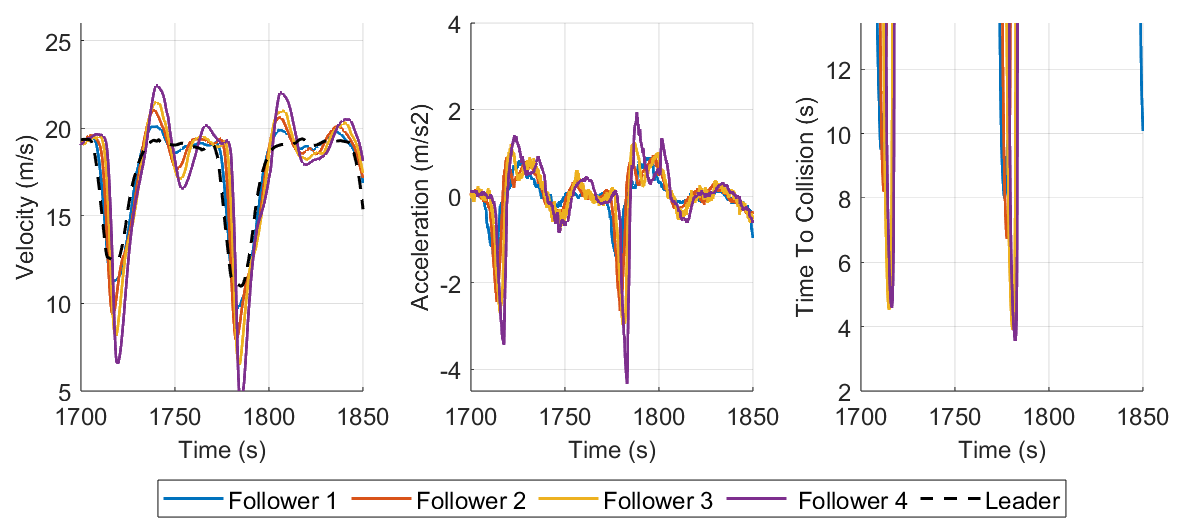}
        \label{fig:OpenACC_AstaZeroMin_Safety_Data}
    }
    \vspace{-2pt}
    \subfloat[Proposed VTG ACC]{
        \includegraphics[width=0.8\textwidth]{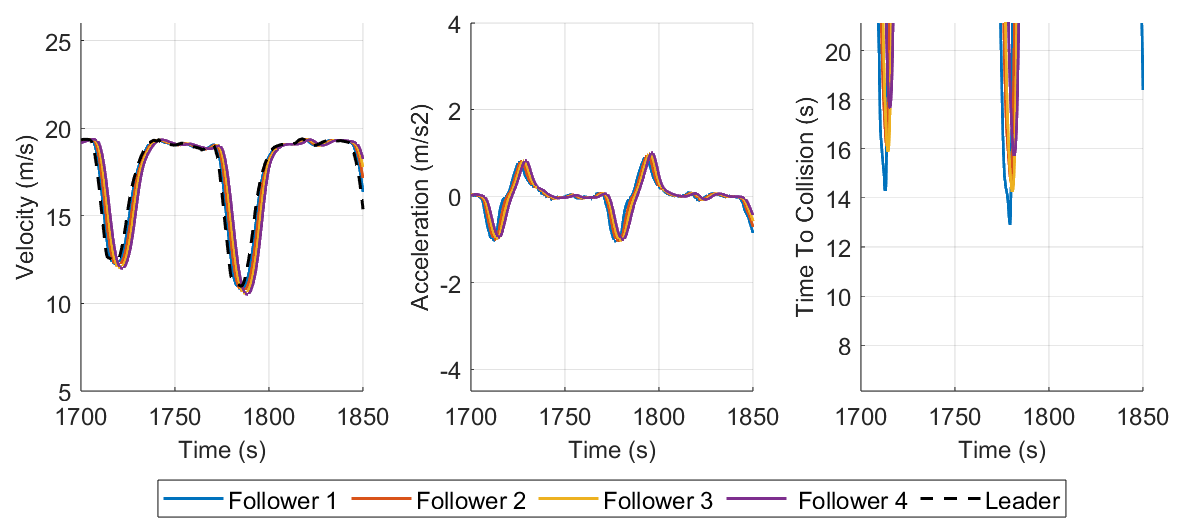}
        \label{fig:OpenACC_AstaZeroMin_Safety_VTG}
    }
    \caption{Simulated trajectories of a platoon of 5 vehicles in a ``dangerous" situation along a straight road using the calibrated CTG ACC and the proposed VTG ACC. Platoon leader velocity is the black dashed line from the AstaZero experimental campaign. All vehicles in this platoon are using ACC with minimum time gap settings.}
    \label{fig:OpenACC_AstaZeroMin_Safety}
\end{figure}

Additionally, Table~\ref{TAB:SSM_OpenACC_AstaZero} outlines more surrogate safety measures to present a rigorous analysis. These measures are the time-exposed TTC and the maximum Deceleration Rate to Avoid Crash (DRAC). The TTC threshold used for the time-exposed TTC computation was $4$s. The significant improvement in these measures brought by the proposed VTG ACC scheme highlight the ability of the feedback VTG policy to react to driving situations that would be deemed dangerous. An example of such ``dangerous" driving situation is extracted from OpenACC dataset and shown in Figure~\ref{fig:OpenACC_AstaZeroMin_Safety}. This situation is selected to exhibit the sharpest deceleration and the least TTC for follower vehicles. As observed, the proposed VTG ACC algorithm is able to handle the situation smoothly, dissipating perturbations and maintaining comfortable and safe accelerations.

\subsection*{Parameter Selection Impact: Numerical Simulations using Field Platoon Data}
\begin{figure}[!ht]
    \centering
    \subfloat[Constant time gap setting of $\tau^\star_i = 1.0$ s]{
        \includegraphics[width=0.9\textwidth]{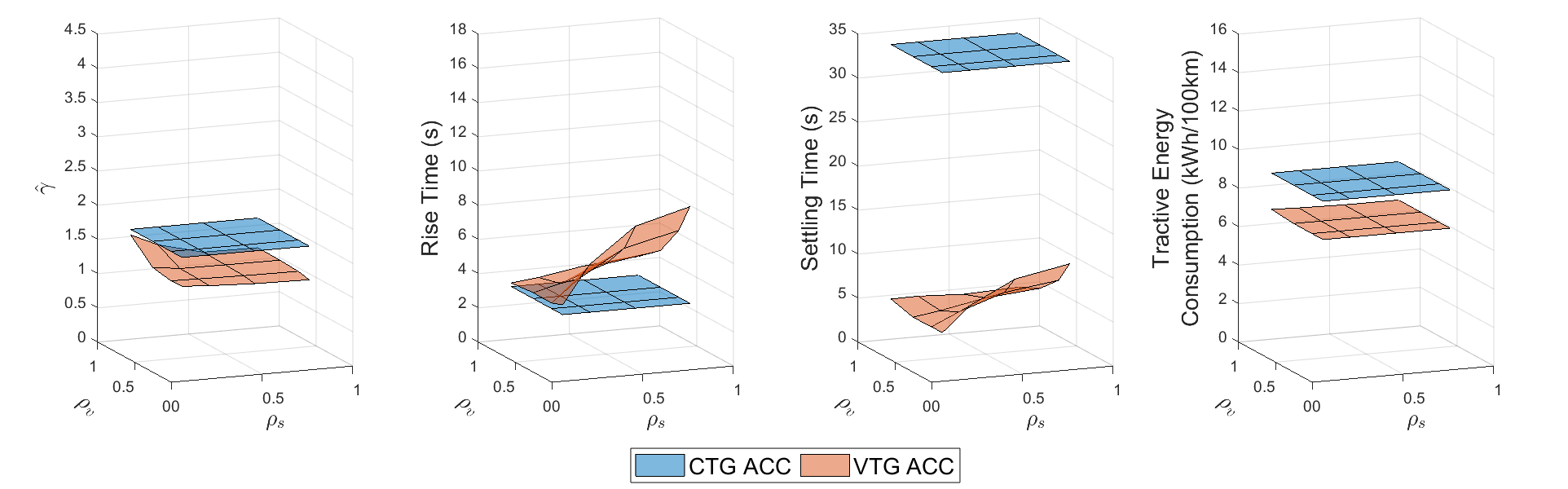}
        \label{fig:UserNeeds_rhoVaried_Tg1}
    }
    \vspace{-2pt}
    \subfloat[Constant time gap setting of $\tau^\star_i = 1.5$ s]{
        \includegraphics[width=0.9\textwidth]{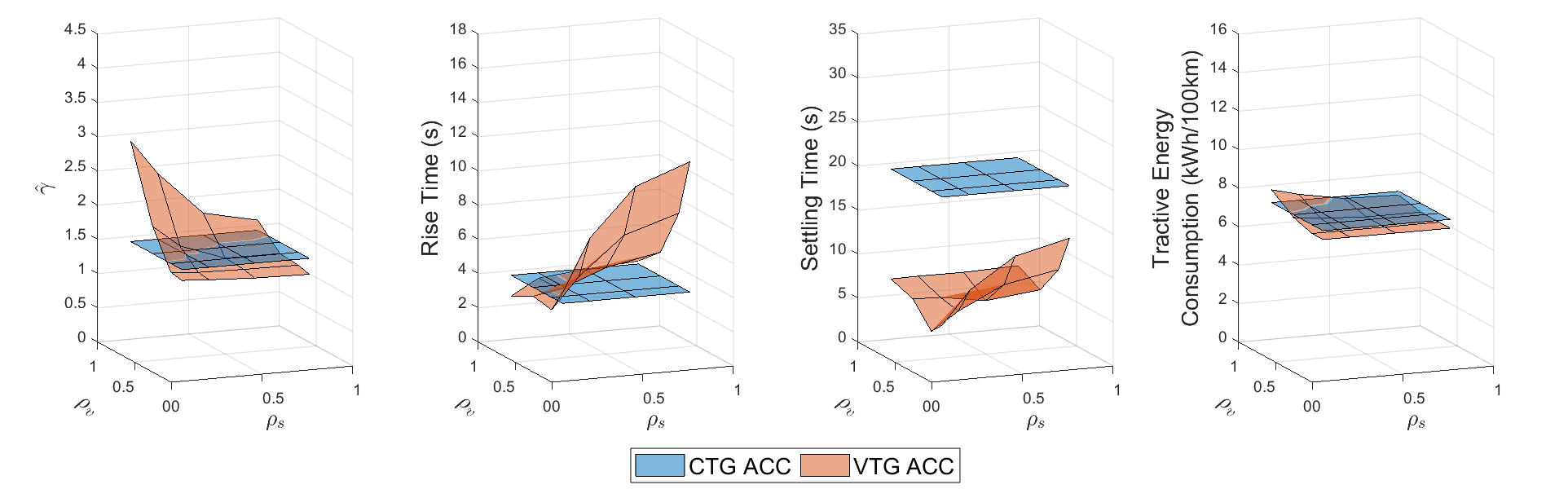}
        \label{fig:UserNeeds_rhoVaried_Tg1p5}
    }
    \vspace{-2pt}
    \subfloat[Constant time gap setting of $\tau^\star_i = 2.0$ s]{
        \includegraphics[width=0.9\textwidth]{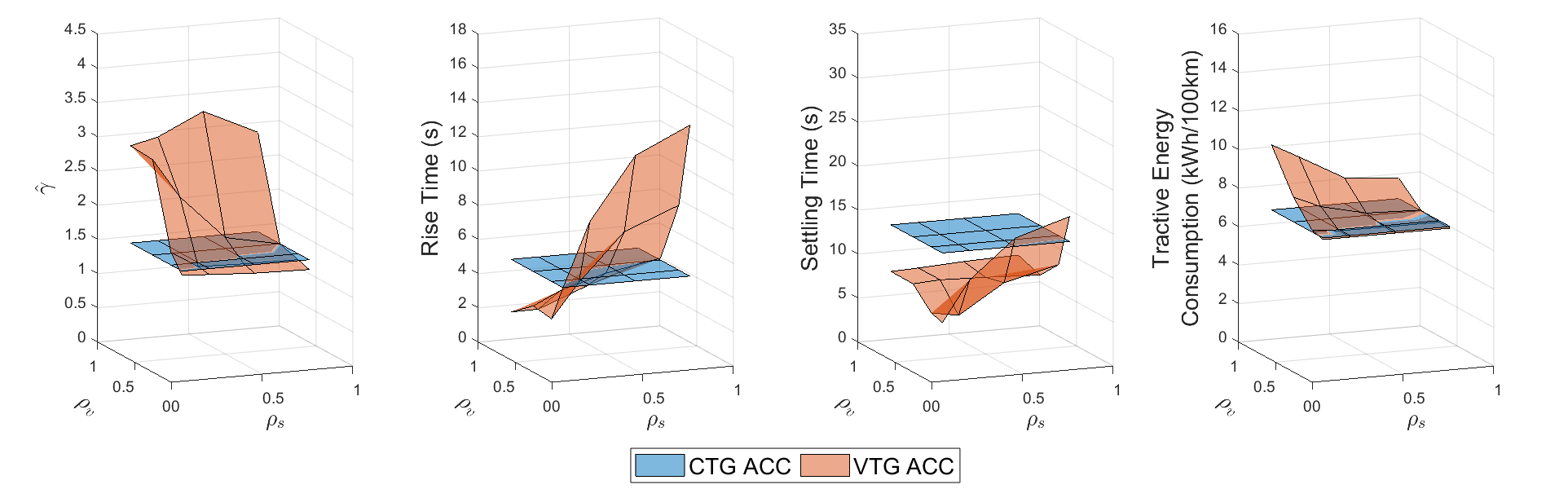}
        \label{fig:UserNeeds_rhoVaried_Tg2}
    }
    \vspace{-2pt}
    \subfloat[Constant time gap setting of $\tau^\star_i = 3.0$ s]{
        \includegraphics[width=0.9\textwidth]{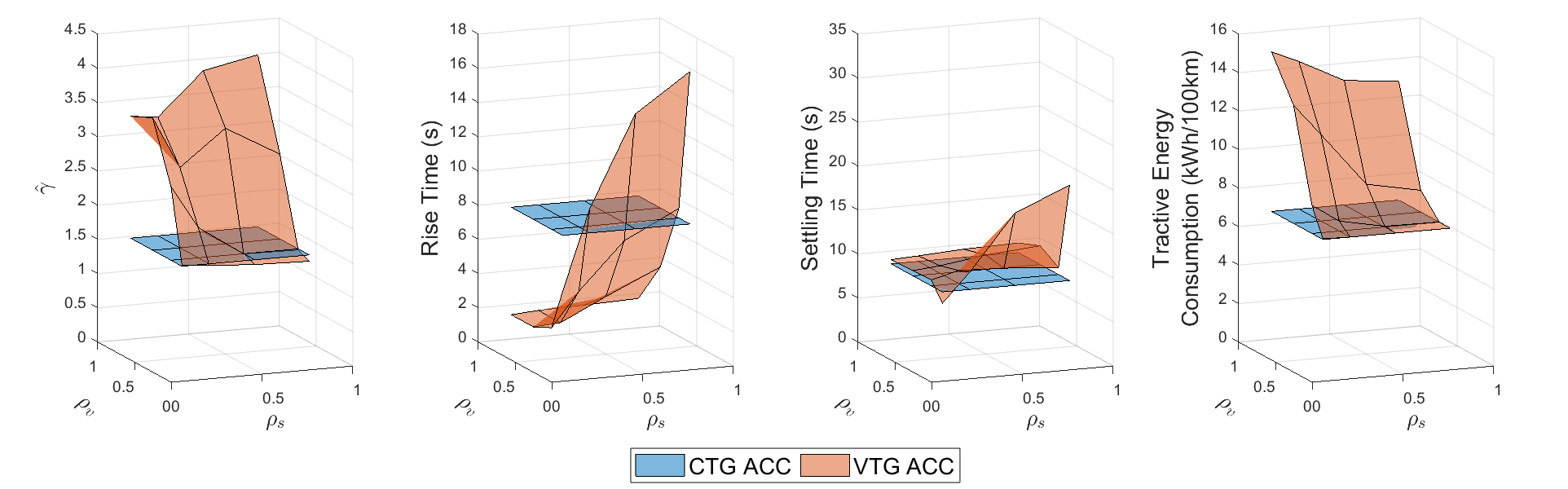}
        \label{fig:UserNeeds_rhoVaried_Tg3}
    }
    \vspace{-2pt}
    \caption{Average performance indicators for a platoon of 5 vehicles along a straight road using the CTG ACC and the proposed VTG ACC after varying the penalty weights, $\rho_s$ and $\rho_v$, at different constant time gap settings $\tau^\star_i$.}
    \label{fig:UserNeeds_rhoVaried}
\end{figure}

In order to assess the effect of the penalty weights, $\rho_s$ and $\rho_v$, a sensitivity analysis, similar to Section~\ref{SUBSEC:UserNeeds}, is carried out in Figure~\ref{fig:UserNeeds_rhoVaried}. At low constant time gap settings, i.e.\ $\tau^\star_i = 1$ s and $\tau^\star_i = 1.5$ s, the VTG ACC controller shows robust performance for various penalty weights compared to the CTG ACC scheme in terms of string stability and energy efficiency. Also, for $\rho_v \leq \rho_s$, the rise time and settling time increase showing less aggressive maneuvers. It becomes critical to select appropriate penalty weights, $\rho_s$ and $\rho_v$, at higher constant time gap values, e.g.\ $\tau^\star_i = 2$ s and $\tau^\star_i = 3$ s; where the potential benefits of the VTG strategy could diminish compared to a CTG strategy.

\subsection*{Cut-in Events Impact: Numerical Simulations using Field Platoon Data}
\begin{figure}[!ht]
    \centering
    \subfloat[Calibrated CTG ACC]{
        \includegraphics[width=0.98\textwidth]{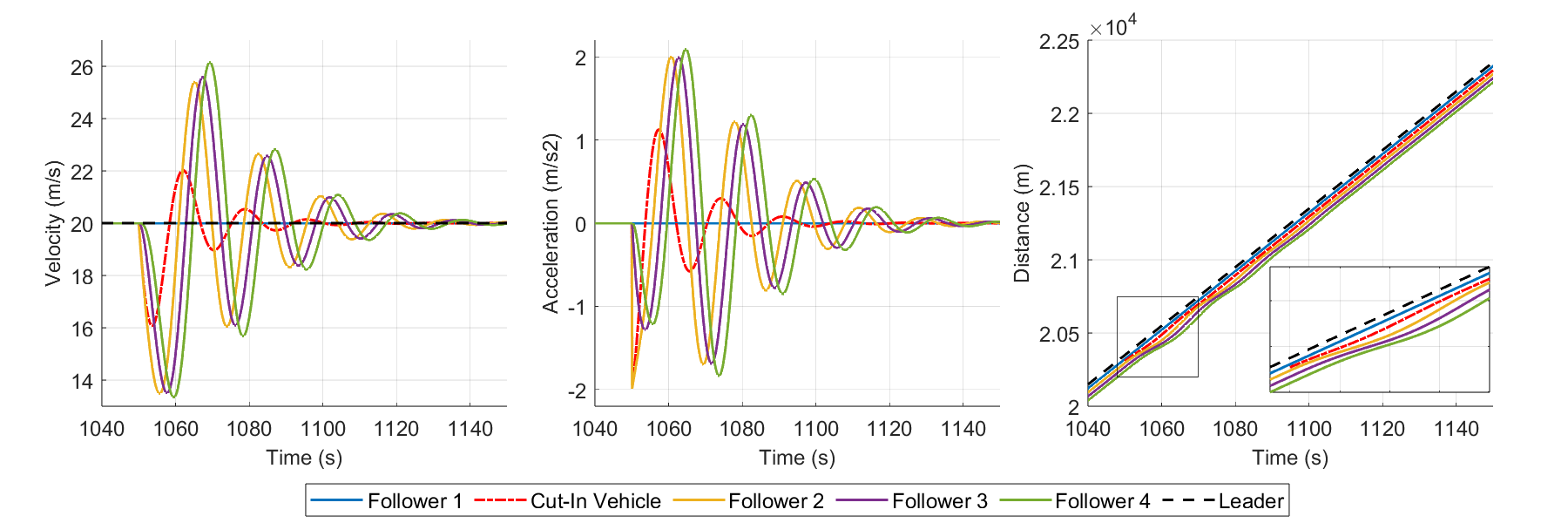}
        \label{fig:CutinEvents_CTG}
    }
    \vspace{-2pt}
    \subfloat[Proposed VTG ACC]{
        \includegraphics[width=0.98\textwidth]{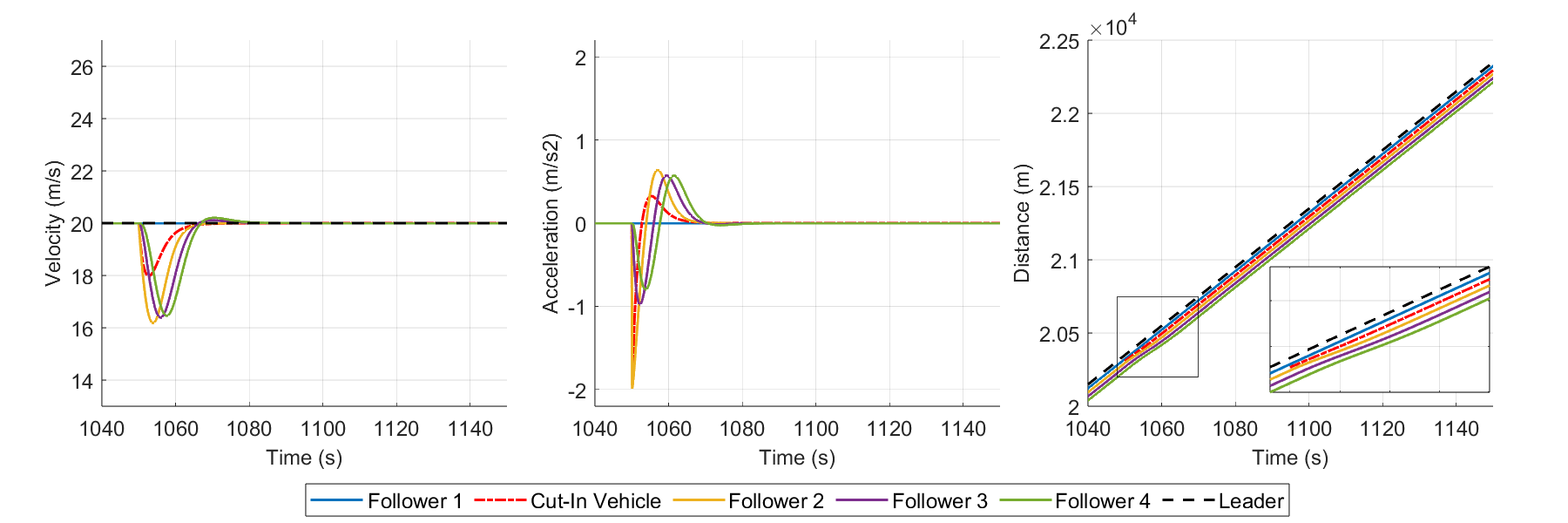}
        \label{fig:CutinEvents_VTG}
    }
    \caption{Simulated trajectories of a platoon of 5 vehicles along a straight road using the calibrated CTG ACC and the proposed VTG ACC. Platoon leader velocity is simulated artificially traveling at a constant speed of $20$ m/s with a cut-in event at $t = 1050$ s. All vehicles in this platoon are using ACC with minimum time gap settings.}
    \label{fig:CutinEvents}
\end{figure}

Here, we assess the effect of cut-in vehicles due to lane changing on the platoon. To this end, the same setup as Section~\ref{SUBSEC:AstaZero} is used; however, an artificial lead vehicle is simulated driving at a constant speed of $20$ m/s and a cut-in vehicle is inserted at $t = 1050$ s. The cut-in vehicle has an initial speed of $20$ m/s, similar to the platoon equilibrium speed to simulate a cut-in event due to lane-changing~\cite{Bang2018}. It is also inserted in the middle of the second and third existing vehicles. Figure~\ref{fig:CutinEvents} shows the response of the calibrated CTG ACC and the proposed VTG ACC controllers. The proposed VTG ACC scheme provides a much more robust response to the cut-in vehicle than the baseline CTG ACC scheme. This shows the reactive variable time gap structure's effect on absorbing and dissipating various disturbances. In addition, the CTG ACC scheme takes much longer to dissipate the disturbances and restore platoon stability than the proposed VTG ACC counterpart.

Also, Table~\ref{TAB:CutinEvents_SSM} presents the surrogate safety measures with a TTC threshold of 5 s used for the time-exposed TTC computation. A lesser maximum DRAC is needed by the VTG ACC controller than that by the CTG one; thus, showcasing the smoothness of the longitudinal maneuver with the ability to preserve safety and string stability. The penalty weights used for this set of simulations were $\rho_s = 0$, $\rho_v = 0.05$, $\rho_u = 0.3$ and $\gamma = 1.0$. Notably, decreasing the penalty weights offers more flexibility to the VTG ACC scheme to absorb the disturbance smoothly and comfortably. This relates to the fact that the penalty weights penalize deviations of the ego vehicle's space headway and velocity away from its equilibrium profile as defined by its constant time gap component. Therefore, higher values of the penalty weights would result in quick dissipation of the disturbance, with respect to the deviations of space headway and ego vehicle's velocity, in an aggressive maneuver.
\begin{table}[!ht]
\centering
\caption{Surrogate safety measures for followers in a platoon of 5 vehicles along a straight road. Platoon leader velocity is simulated artificially traveling at a constant speed of $20$ m/s with a cut-in event at $t = 1050$ s. All vehicles in this platoon are using ACC with minimum time gap settings.}
\label{TAB:CutinEvents_SSM}
\begin{tabular}{l|c|c|c}
                & \begin{tabular}[c]{@{}c@{}}Minimum TTC\\ (s)\end{tabular} & \begin{tabular}[c]{@{}c@{}}Time-Exposed TTC\\ (s)\end{tabular} & \begin{tabular}[c]{@{}c@{}}Maximum DRAC\\ ($\text{m/s}^2$)\end{tabular} \\ \hline
CTG ACC         & 4.28                                                      & 2.7                                                           & 0.51                                                                    \\
VTG ACC         & 11.56                                                     & 0                                                              & 0.11                                                                    \\ \hline
\end{tabular}
\end{table}

%%%%%%%%%%%%%%%%%%%%%%%%%%%%%%%%%%%%%%%%%%%%%%%%%%%%%%%%%%%%%%%%%%%%%%%%%%%%%%%%%%%%%%%%%%%
%%%%%%%%%%%%%%%%%%%%%%%%%%%%%%%%%%%%%%%%%%%%%%%%%%%%%%%%%%%%%%%%%%%%%%%%%%%%%%%%%%%%%%%%%%%
%\newpage
\section*{Appendix C} \label{SEC:AppendixC}
In this section, we present a parameter tuning method to optimize the penalty weights, $\rho_s, \rho_v, \rho_u$ and $\gamma$ of the proposed VTG ACC algorithm. Here, we use the same setup as Section~\ref{SUBSEC:AstaZero}; where one set of field platoon data from the AstaZero campaign is used for the optimization procedure and another is used for testing the optimized parameters. The parameter tuning optimization procedure is expressed as follows.
\begin{subequations}
\begin{align}
    \hat{\Omega} &= \min_{\Omega \in [0,1]^4 } \frac{1}{N} \sum_{i=1}^{N} \hat{\gamma}_i + TET_i + E_i \\
    & \Omega = \begin{bmatrix} \rho_s & \rho_v & \rho_u & \gamma \end{bmatrix}^T \\
    & TET_i = \sum_{t=1}^{T} \delta T_s, \qquad \delta = \begin{cases} 
1 \quad \text{if }\,  0 < TTC(t) < TTC^\star \\ 0 \quad \text{otherwise} \end{cases}
\end{align}
\end{subequations}
where $TET_i$ is the time-exposed TTC of vehicle $i$, $TTC^\star = 4$ s is the threshold TTC to deem a time instant as an exposed or possibly unsafe instant, $T_s = 0.1$ s is the sampling time interval, and $E_i$ is the tractive energy consumption of vehicle $i$ computed by \eqref{EQ:Ec}. Also, $N$ denotes the number of platoon followers. 

\begin{figure}[!ht]
    \centering
    \subfloat[Response for delay-free case]{
        \includegraphics[width=0.8\textwidth]{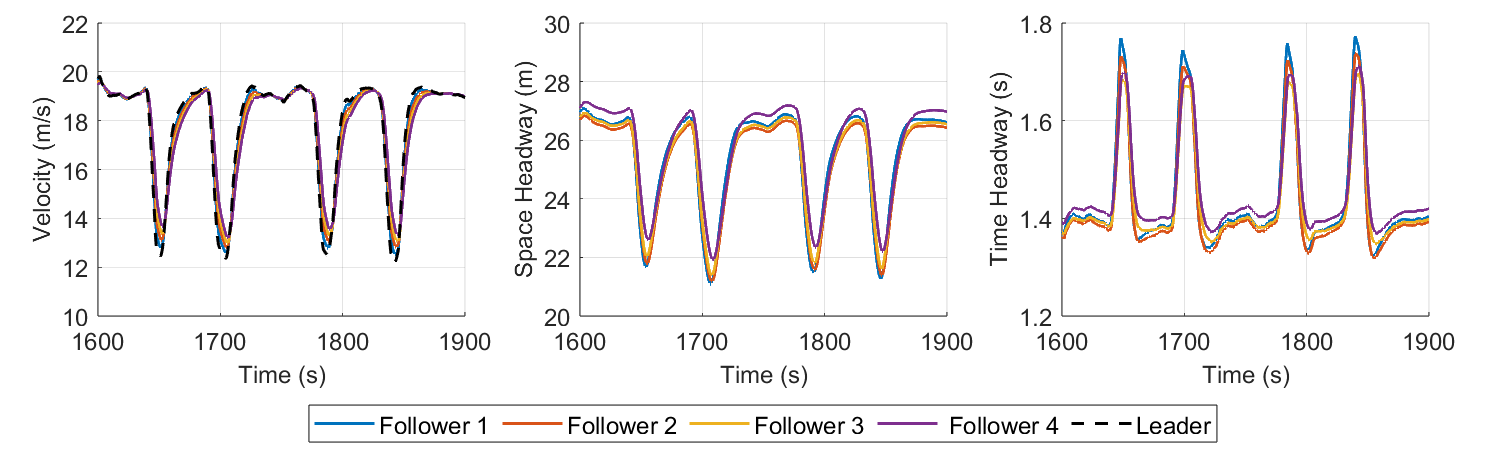}
        \label{fig:AstaZero_Tuning_VTG}
    }
    \vspace{-2pt}
    \subfloat[Performance Indicators for delay-free case]{
        \includegraphics[width=0.8\textwidth]{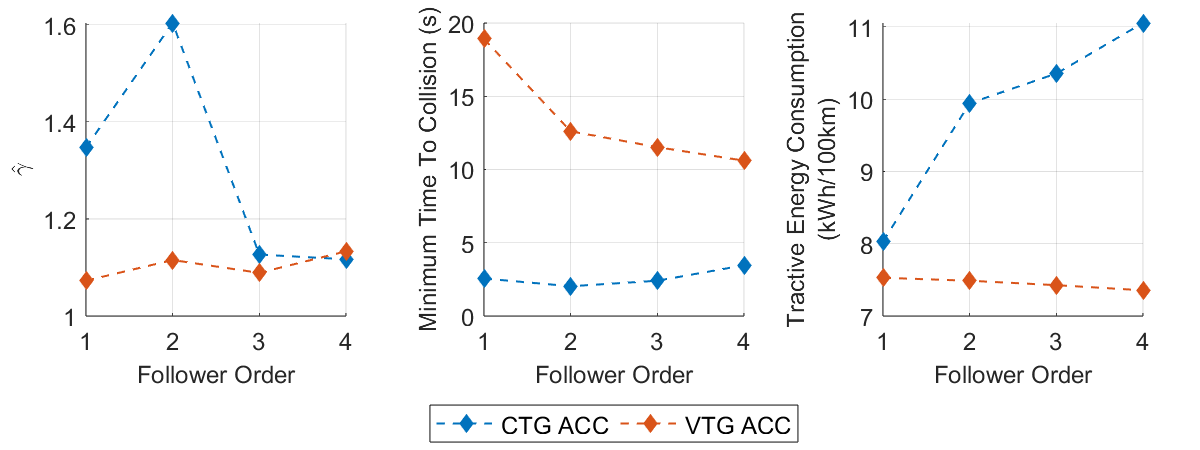}
        \label{fig:AstaZero_Tuning_Perf}
    }
    \vspace{-2pt}
    \subfloat[Response for delay-prone case]{
        \includegraphics[width=0.8\textwidth]{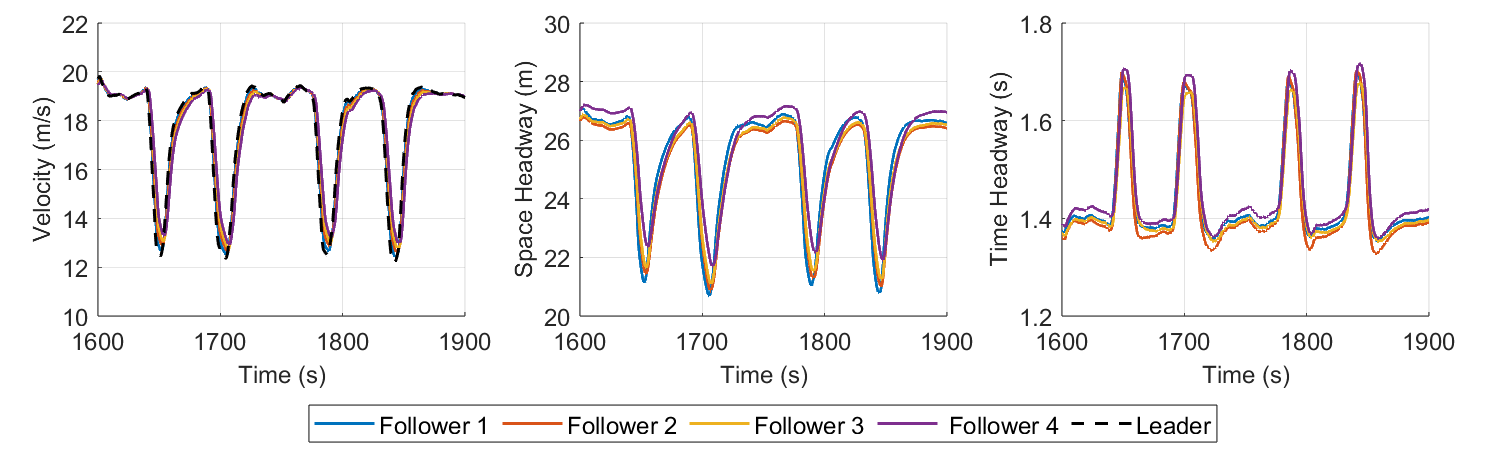}
        \label{fig:AstaZero_Tuning_Delay_VTG}
    }
    \vspace{-2pt}
    \subfloat[Performance Indicators for delay-prone case]{
        \includegraphics[width=0.8\textwidth]{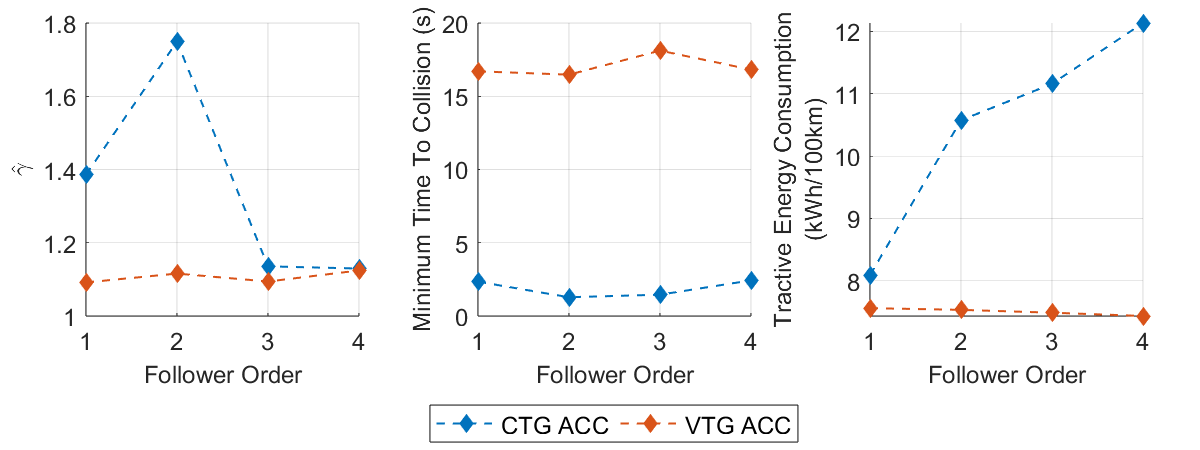}
        \label{fig:AstaZero_Tuning_Delay_Perf}
    }
    \vspace{-2pt}
    \caption{Simulation responses and average performance indicators for a platoon of 5 vehicles along a straight road using the CTG ACC and the proposed VTG ACC using the optimal set of penalty weights, $\rho_s, \rho_v, \rho_u$ and $\gamma$, for both the delay-free and delay-prone cases.}
\end{figure}

This optimization problem is solved using a differential evolution algorithm with a population size of 20 and a maximum of 50 iterations. The optimal set of parameters for the delay-free case (i.e.\ without considering low-level vehicle lag dynamics) is $\Omega^\star = \begin{bmatrix} 0.1 & 0.73 & 0.56 & 1.0 \end{bmatrix}^T$. The simulation results via the testing campaign are shown in Figure~\ref{fig:AstaZero_Tuning_VTG} and Figure~\ref{fig:AstaZero_Tuning_Perf}. If we consider the delay-prone case with a lag constant of $\tau_a = 0.1$ s and an input delay of $\tau_D = 0.2$ s, the optimal set of parameters is $\Omega^\star = \begin{bmatrix} 0.01 & 1.0 & 0.95 & 0.92 \end{bmatrix}^T$. The simulation results via the testing campaign are shown in Figure~\ref{fig:AstaZero_Tuning_Delay_VTG} and Figure~\ref{fig:AstaZero_Tuning_Delay_Perf}. As observed, the optimized parameters perform well and show optimized indicators compared to the baseline CTG ACC counterpart. Therefore, this general workflow can be utilized to automatically tune the proposed VTG ACC parameters for any specific use case. 

\clearpage

%Bibliography
\bibliographystyle{ieeetr}  
\bibliography{references}

\end{document}